\documentclass[prd,showpacs,amsmath,showkeys,twocolumn,floatfix,amssymb, preprintnumbers, nofootinbib, superscriptaddress]{revtex4} 
\usepackage{epsfig,dcolumn}
\usepackage{graphicx}
\usepackage{comment} 
\usepackage[usenames]{color}
\usepackage{graphicx}
\usepackage{bm}
 \usepackage{ifpdf}

\def\lsim{\mathrel{\rlap{\lower4pt\hbox{\hskip1pt$\sim$}}
    \raise1pt\hbox{$<$}}}
\def\gsim{\mathrel{\rlap{\lower4pt\hbox{\hskip1pt$\sim$}}
    \raise1pt\hbox{$>$}}}
\begin{document}

\title{  One spatial dimensional finite volume three-body interaction for a short-range potential}

\author{Peng~Guo}
\email{pguo@jlab.org}

\affiliation{Department of Physics and Engineering,  California State University, Bakersfield, CA 93311, USA.}

\date{\today}

\begin{abstract} 

In this work, we use McGuire's model  to describe scattering of three spinless identical particles   in one spatial dimension, we first present    analytic solutions of Faddeev's equation  for scattering  of three   spinless   particles in free space. The three particles interaction in finite volume  is derived subsequently, and the quantization conditions by matching wave functions in free space and finite volume are presented  in terms of two-body scattering phase shifts.   The quantization conditions   obtained in this work  for short range interaction are  L\"uscher's formula like and   consistent with Yang's results   in \cite{Yang:1967bm}.

  \end{abstract} 

\pacs{ }

\maketitle

\section{Introduction}

Three-particle interaction plays an important role in hadron physics. In certain reaction processes, three-particle dynamics may be a crucial component of reaction. For a example, the discrepancy of    decay width of \mbox{$\eta \rightarrow 3 \pi$} between experimental measurement \cite{PDG-2012} and $\chi$PT  calculations \cite{Cronin:1967jq,Osborn:1970nn,Gasser:1984pr,Bijnens:2007pr} can only be well-understood when three-body dynamics are properly considered \cite{Kambor:1995yc,Anisovich:1996tx,Colangelo:2009db,Lanz:2013ku,Schneider:2010hs,Kampf:2011wr,Guo:2015zqa}. In the past, many different approaches have been developed  to describe three-body dynamics, for instance, quantum field theory based relativistic Bethe-Salpeter equations \cite{Taylor:1966zza,Basdevant:1966zzb,Gross:1982ny}, non-relativistic Faddeev's equation \cite{Faddeev:1960su,Faddeev:1965}, and dispersion relation oriented Khuri-Treiman equation \cite{Khuri:1960zz,Bronzan:1963xn,Aitchison:1965kt,Aitchison:1965zz,Aitchison:1966kt,Pasquier:1968zz,Pasquier:1969dt,Guo:2014vya,Guo:2014mpp,Danilkin:2014cra,Guo:2015kla}. Unfortunately, either approach provides a non-expert friendly framework due to sophistication of three-body dynamics. In recent years, three-body dynamics start regaining some popularities in hadron physics community for many reasons. For examples,   precision theoretical hadron-interaction framework is urgently needed for data analysis when high statistic data become available, and a sensible finite volume theory of three-body interaction is also currently demanded by lattice QCD community.

In lattice QCD calculation, because computation is performed in Euclidean space, we do not have direct access to scattering amplitudes \cite{Maiani:1990ca}. Fortunately, taking advantage of periodic boundary condition, a relation between the energy spectrum extracted from lattice QCD computation  and     two-body  scattering amplitudes  in elastic region is established \cite{Luscher:1990ux}, which is usually referred to L\"uscher's formula. The extension of framework to moving frames and to inelastic channels have also been developed by many authors \cite{Rummukainen:1995vs,Lin:2001ek,Christ:2005gi,Bernard:2007cm,Bernard:2008ax,He:2005ey,Lage:2009zv,Doring:2011vk,Aoki:2011gt,Briceno:2012yi,Hansen:2012tf,Guo:2012hv}.  The  finite volume scattering formalism  has been   proven valid and effective in   lattice community for extracting   hadron-hadron two-body scattering information \cite{Aoki:2007rd,Sasaki:2008sv,Feng:2010es,Dudek:2010ew,Beane:2011sc,Lang:2011mn,Aoki:2011yj,Dudek:2012gj,Dudek:2012xn,Wilson:2014cna,Wilson:2015dqa,Dudek:2016cru}.

There have been some attempts on finite volume three-body interactions in recent years \cite{Kreuzer:2008bi,Kreuzer:2009jp,Kreuzer:2012sr,Polejaeva:2012ut,Briceno:2012rv,Hansen:2014eka,Hansen:2015zga,Hansen:2016fzj}. These  recent   developments for  finite volume three-body scattering problem  \cite{Kreuzer:2008bi,Kreuzer:2009jp,Kreuzer:2012sr,Polejaeva:2012ut,Briceno:2012rv,Hansen:2014eka,Hansen:2015zga,Hansen:2016fzj}   are typically momentum   representation of  quantum field theory approaches, diagrammatic approaches or Faddeev equation  based method.  None of these developments are mathematics and physics friendly  to majority of people in physics community. Hence, it is fair and reasonable  to raise the question that  how one would check all these over sophisticated quantization conditions of finite volume three-body problem presented in these works? In present work, we aim to find a simple and exactly solvable three-body problem, so that the analytic results of quantization conditions in this simple case can be found. The result may serve as a calibration to more realistic treatments of three-body problem in different approaches \cite{Kreuzer:2008bi,Kreuzer:2009jp,Kreuzer:2012sr,Polejaeva:2012ut,Briceno:2012rv,Hansen:2014eka,Hansen:2015zga,Hansen:2016fzj}. 
 In order to make the three-body scattering problem as simple as possible and exactly solvable, we will ignore the relativistic effect and also constrain ourself to one spatial dimension, so that    analytic solutions can be found and  infinite sum in finite volume  are easily carried out. We will further consider three non-relativistic particles with equal masses, and the pair-wise and short range  interactions among three particles.  Under above mentioned assumptions, a simplified Faddeev's equation in free space  is established and solved analytically. Instead of attacking finite volume problem in momentum representation, we employ the approach developed in \cite{Guo:2012hv,Guo:2013vsa},  and work our way out with wave function of three-body in configuration representation. In our approach of solving three-body problem in finite volume, there are three basic steps: first of all, the solutions of Faddeev's equation  are used to construct free space three-body wave function.   Next, for the problem of  three particles   in a finite box, we show that the finite volume wave function of three-body interaction can be constructed through free space three-body wave function. At last, the matching of finite volume wave function and free space wave function yields the Lusher's formula like  quantization conditions of three-body interaction, which are sets of relations between two-body phase shift and three-particle momenta in a finite box.
\begin{align}
& \cot (\frac{P}{3} + \frac{q_{3}}{2})L + \cot  \left (  \delta (-q_{31})-\delta (-q_{23})  \right ) =0 , \nonumber \\
&  \cot (\frac{P}{3} + \frac{q_{1}}{2})L + \cot  \left ( - \delta (-q_{31}) - \delta (q_{12}) \right ) =0, \nonumber \\
&  \cot (\frac{P}{3} + \frac{q_{2}}{2})L + \cot  \left ( \delta (-q_{23}) + \delta (q_{12}) \right ) =0 ,\nonumber
\end{align}
where $\delta$ denotes the two-body scattering phase shift and $L$ refers to the size of box. The total momentum, $P$,   relative momentum $q_{ij}$ between i-th and j-th particles,  and relative momenta $q_{k}$ between k-th particle and pair $(ij)$ will be explained in Section \ref{3bfree}.

The advantage of using wave function in configuration representation is that   first of all, the wave function approach is   close to the way of solving traditional  quantum mechanics problems in a periodic potential. A short review of our formalism for two-body interaction in finite volume and its applications to  exactly solvable quantum mechanical models are listed in Appendix \ref{2bscatt}.  Secondly,   also the most important fact is that  the asymptotic form of  wave function  displays  physical on-shell transition amplitudes. As a well-known fact,    the solutions of Faddeev's equation are not equivalent to physical transition amplitudes \cite{Faddeev:1965,Gerjuoy:1971,Osborn:1970zz,Osborn:1973zz}. Three-body scattering amplitude posses singularities of poles and $\delta$-functions, the physical transition amplitudes are in fact associated to the residue functions of these singularities. These singularities are the consequence of existence of different distinct physical processes in three-body system. For examples, unlike in two-body system, the formation of a bound pair are not precluded by energy conservation, a pole is thus introduced by the presence of a two-body bound state. 
The singularity of $\delta$-functions are associated with disconnected diagrams with  a unscattered third particle.  Because of  these singularities in three-body amplitude,  the three-body wave function in configuration representation may have   several  different pieces that decrease at different rate  and describe different   physical processes.  The physical transition amplitudes for different physical processes are thus also defined by   asymptotic forms of three-body wave function \cite{Nuttall:1969mp,Newton:1972xc,Merkurev:1976uy,Osborn:1970zz,Osborn:1973zz}.

In this work, to describe the scattering of three-body system,  we adopt a one dimensional model with the interaction of equal strength, pair-wise $\delta$-function potential  that was developed by McGuire in  \cite{McGuire:1964zt}. A brief review of McGuire's model is provided in Appendix \ref{McGuire}. The McGuire's model is physically simple, but can still provide us a qualitative description of three-body scattering process in finite volume. McGuire's model was originally solved by ray-tracing and geometric optics consideration method  \cite{McGuire:1964zt}, McGuire found that diffraction effects in this particular model are all cancelled out,  thus   no new momenta are created over scattering process, though momenta are allowed to be rearranged among three particles in final state. In consequence, any dissociation or recombination of particles out of or into bound states are forbidden, bound states sectors are decoupled from three free particles sector. Breakup process never happens when a particle is incident on a bound state \cite{Dodd:1970,Majumdar:1972,Gerjuoy:1985}. Nevertheless, McGuire's model still encompass rearrangement effects among three particles, it may even represent  more realistic physical models of     short range interaction. Although, the quantization conditions for  three-body problem in finite volume are obtained by considering a particular model, the final results are presented in terms of two-body phase shifts. These quantization conditions presented in this way may still be valid for a more general pair-wise short range potential.  Moreover, the quantization conditions may be tested numerically in future by one dimensional lattice models, such as  ones developed in \cite{Gattringer:1992np,Guo:2013vsa}.

The paper is organized as follows.   In Section \ref{3bfree} we discuss the free space three-particle system.   The finite volume three-particle system is presented in Section \ref{3bbox}. The summary and outlook are given in Section \ref{summary}.

\section{Three-body scattering for short range interaction in free space}\label{3bfree}

Considering three spinless identical particles scattering, the short range interactions among three-particle are   pair-wise and equal strength for all pairs, \mbox{$V(r)=V_{0}\delta(r)$}. In general,  the kernel for Faddeev's integral equation is off-shell two-body scattering amplitudes, off-shell kernel usually complicates three-body integral equations even in one dimension and   causes   difficulties of solving Faddeev's equation.  Fortunately, for short range $\delta$-function potential, two-body scattering amplitude appears completely on energy shell, see Eq.(\ref{deltatamp}). This feature dramatically simplifies Faddeev's integral equation, so that finding an analytic solution is possible. For completeness, a brief review of formal scattering theory and general framework of Faddeev's equation is presented in Appendix \ref{formalscatt}.

 The wave function of scattering three-particle satisfies Schr\"odinger  equation,
\begin{align}
& \left [- \frac{1}{2 m} \sum_{i=1}^{3} \frac{d^{2}}{d x_{i}^{2}}  + V_{0}\delta(r_{12}) + V_{0} \delta(r_{23}) + V_{0}\delta(r_{31})  - E \right ]  \nonumber \\
&\quad \quad  \times \Psi(x_{1}, x_{2}, x_{3}; p_{1},p_{2},p_{3})=0,
\end{align}
where the mass of particle is $m$,  the total energy of three-particle system is \mbox{$E= \sum_{i=1}^{3}\frac{p_{i}^{2}}{2m}$}, where $p_{i}$ (\mbox{$i=1,2,3$})  stands for the particle's momenta in initial state. The center of mass  position is given  by \mbox{$R= \frac{ x_{1}+ x_{2}+x_{3}}{3}$}. \mbox{$r_{ij}=x_{i}-x_{j}$} refer to relative position between i-th and j-th particles,  and \mbox{$r_{k}=\frac{x_{i} + x_{j}}{2} - x_{k}$}  denote the relative position between k-th particle and pair $(ij)$.  The conjugate total and relative momenta  are given by  \mbox{$P = p_{1} + p_{2}+p_{3}$}, \mbox{$q_{ij} = \frac{p_{i}-p_{j}}{2}$} and \mbox{$q_{k} = \frac{p_{i} + p_{j} - 2 p_{k}}{3}$} respectively.  A change of the pair of relative coordinates and corresponding conjugate momenta from $(ij)k$ configuration to other configurations, {\it e.g.} $(jk)i$ configuration in which relative coordinates and conjugate momenta are expressed in terms of \mbox{$(r_{jk}, r_{i})$} and \mbox{$(q_{jk}, q_{i})$}, is accomplished by 
\begin{align}
& r_{jk} = - \frac{1}{2} r_{ij} + r_{k}, \ \ \ \ r_{i} = -\frac{3}{4} r_{ij} - \frac{1}{2} r_{k}, \nonumber \\
& q_{jk} = - \frac{1}{2} q_{ij} + \frac{3}{4} q_{k}, \ \ \ \ q_{i} = - q_{ij} - \frac{1}{2} q_{k}, \label{momentarel}
\end{align} 
where $(ij)k$ or $(jk)i$ always follows cyclic permutation of $(1,2,3)$.

 The total wave function of three particles can be expressed by the product of a plane wave, $e^{ i P R} $, which describes center of mass motion, and the relative wave function, \mbox{$\psi(r_{ij}, r_{k};q_{ij}, q_{k})$}, which   describes relative motions of three particles, \mbox{$\Psi(x_{1}, x_{2},x_{3}; p_{1},p_{2},p_{3}) = e^{ i P R} \psi(r_{ij}, r_{k};q_{ij}, q_{k})$}.  
 For three free particles scattering, the wave function has the following form \cite{Faddeev:1960su,Faddeev:1965},
  \mbox{$\Psi=\Psi_{(0)} +  \sum_{\gamma=1}^{3} \Psi_{(\gamma)}$},  
 where \mbox{$\Psi_{(0)}$} refers to the incoming free wave, and $\Psi_{(\gamma)}$ satisfies equation,
\begin{align}
& \left [- \frac{1}{2 m} \sum_{i=1}^{3} \frac{d^{2}}{d x_{i}^{2}}  + V_{0} \delta(r_{\alpha \beta})   - E \right ] \Psi_{(\gamma)}  \nonumber \\
&\quad   \quad \quad     = -V_{0}\delta(r_{\alpha \beta})  \left [ \Psi_{(0)}+ \Psi_{(\alpha)} + \Psi_{(\beta)}  \right ] , \ \ \ \gamma \neq \alpha \neq \beta. \label{3bpsik}
\end{align}
The integral representation of Eq.(\ref{3bpsik}) for relative wave function, \mbox{$ \psi_{(\gamma)} (r_{\alpha\beta}, r_{\gamma}; q_{ij}, q_{k}) =e^{-i P R} \Psi_{(\gamma)} (x_{1}, x_{2}, x_{3}; p_{1},p_{2},p_{3})$}, is given by
\begin{align}
& \psi_{(\gamma)}  (r_{\alpha \beta}, r_{\gamma};q_{ij}, q_{k}) \nonumber \\
&     = \int_{-\infty}^{\infty}  d r'_{\alpha \beta} d r'_{\gamma}  G_{(\gamma)} (r_{\alpha \beta}- r'_{\alpha \beta}, r_{\gamma}- r'_{\gamma}; z_{\sigma} )  \nonumber \\
&         \times  m V_{0} \delta(r'_{\alpha \beta}) \left [ \psi_{(0)}(r'_{\alpha \beta}, r'_{\gamma};q_{ij}, q_{k})   + \psi_{(\alpha)}  (r'_{\beta\gamma}, r'_{\alpha};q_{ij}, q_{k}) \right.  \nonumber \\
& \quad \quad  \quad \quad \quad \quad    \quad  \quad    \quad          \left.  +  \psi_{(\beta)} (r'_{\gamma \alpha}, r'_{\beta};q_{ij}, q_{k})  \right ]  , \label{3bintpsik}  
\end{align}
where \mbox{$z_{\sigma} = \sigma^{2} + i \epsilon$} and \mbox{$\sigma^{2} = mE - \frac{P^{2}}{6}= q_{ij}^{2} + \frac{3}{4} q^{2}_{k}$} (\mbox{$k=1,2,3$}), and   the Green's function $G_{(\gamma)}$  satisfies equation,
\begin{align}
& \left [  z_{\sigma} + \frac{d^{2}}{d r_{\alpha \beta}^{2}} + \frac{3}{4}  \frac{d^{2}}{d r_{\gamma}^{2}} -m V_{0} \delta (r_{\alpha \beta})    \right ] G_{(\gamma)} (r_{\alpha \beta} -r'_{\alpha \beta} , r_{\gamma} ; z_{\sigma} )  \nonumber \\
&\quad \quad  \quad \quad  \quad \quad  \quad \quad  \quad \quad  \quad \quad    =  \delta(r_{\alpha \beta}- r'_{\alpha \beta})   \delta(r_{\gamma})  ,   \label{Green}
\end{align}
and  the solution of Eq.(\ref{Green}) is given by
\begin{align}
 & G_{(\gamma)} (r_{\alpha \beta} -r'_{\alpha \beta} , r_{\gamma} ; z_{\sigma} )   \nonumber \\
 &   = \int_{-\infty}^{\infty}  \frac{d q'_{\alpha \beta}}{2\pi} \frac{d q'_{\gamma}}{2\pi} \frac{ \sum_{\mathcal{P} = \pm} \psi_{\mathcal{P}} (r_{\alpha \beta}; q'_{\alpha \beta}) \psi^{*}_{\mathcal{P}} (r'_{\alpha \beta}; q'_{\alpha \beta}) e^{i q'_{\gamma} r_{\gamma}}}{ z_{\sigma}  -  {q'}^{2}_{\alpha \beta}     -\frac{3}{4} {q'}^{2}_{\gamma}  }.  
\end{align}
The $\psi_{\pm} (r_{\alpha \beta}, q_{\alpha \beta})$ are parity two-body wave functions of pair $(\alpha \beta)$, and  the solution of $\psi_{\pm}$ for $\delta$-function potential  read
\begin{align}
 \psi_{\mathcal{P}} (r; k) &  = \frac{e^{i k r} + \mathcal{P} e^{-i k r}}{2}  +  i t_{\mathcal{P}} (\sqrt{z_{k}})Y_{\mathcal{P}}(k)  e^{i \sqrt{z_{k}} |r|}    ,  
\end{align} 
where the on-shell two-body scattering amplitudes, $t_{\pm}$,  are given  in Eq.(\ref{deltatamp}): \mbox{$t_{+}(k) =- \frac{ m V_{0}}{2k+i m V_{0}} $} and \mbox{$t_{-}(k)=0$}. $t_{\pm}$ are normalized by relation:
\mbox{$\frac{t_{\mathcal{P} } - t^{*}_{\mathcal{P}} }{2i} = t^{*}_{\mathcal{P}} t_{\mathcal{P}}$}.
Using unitarity relation of two-body amplitude, it can  be shown that
\begin{align}
& \int_{-\infty}^{\infty}   d r'_{\gamma}  G_{(\gamma)} (r_{\alpha \beta}- r'_{\alpha \beta}, r_{\gamma}- r'_{\gamma};  z_{\sigma} )    V_{0} \delta (r'_{\alpha \beta})  e^{i  q r'_{\gamma}} \nonumber \\
&=   \frac{e^{i \sqrt{\sigma^{2}   -\frac{3}{4} q^{2} } | r_{\alpha \beta} | }e^{i  q  r_{\gamma}  } }{2 i \sqrt{\sigma^{2}     -\frac{3}{4} q^{2} }  }      \left [ 1 + i t_{+}(\sqrt{\sigma^{2}     -\frac{3}{4} q^{2} } )   \right ] V_{0}\delta(r'_{\alpha \beta})    . 
\end{align}
Therefore, the Eq.(\ref{3bintpsik}) can be written as
\begin{align}
& \psi_{(\gamma)}  (r_{\alpha \beta}, r_{\gamma};q_{ij}, q_{k})  =  \int_{-\infty}^{\infty}   \frac{d q}{2\pi}   e^{i \sqrt{\sigma^{2}    -\frac{3}{4} q^{2}} | r_{\alpha \beta} | }e^{i q  r_{\gamma}  }   \nonumber \\
&   \times \frac{    \left [ 1 + i t_{+}(\sqrt{\sigma^{2}   -\frac{3}{4} q^{2} } )   \right ] }{ 2 i \sqrt{\sigma^{2}    -\frac{3}{4} q^{2}}  } \int_{-\infty}^{\infty}  d r'_{\alpha \beta} d r'_{\gamma}    e^{ - i q r'_{\gamma} }   m V_{0} \delta(r'_{\alpha \beta}) \nonumber \\
&         \times \left [ \psi_{(0)}(r'_{\alpha \beta}, r'_{\gamma};q_{ij}, q_{k})   + \psi_{(\alpha)}  (r'_{\beta\gamma}, r'_{\alpha};q_{ij}, q_{k}) \right.  \nonumber \\
& \quad \quad  \quad \quad \quad      \quad  \quad    \quad       \quad    \quad     \left.  +  \psi_{(\beta)} (r'_{\gamma \alpha}, r'_{\beta};q_{ij}, q_{k})  \right ]    . \label{wavegamma}
\end{align}

Next, let's   introduce   amplitudes, $T_{(\gamma)}$, by
\begin{align}
T_{(\gamma)} (k ; & q_{ij}, q_{k}) =- \int_{-\infty}^{\infty}  d r_{\alpha \beta} d r_{\gamma}  e^{-i k  r_{\gamma}} m V_{0} \delta (r_{\alpha \beta}) \nonumber \\
& \quad \times   \psi ( r_{\alpha \beta}, r_{\gamma} ; q_{ij}, q_{k}), \ \ \ \  \alpha \neq \beta \neq \gamma . \label{Tgammawav}
\end{align}
Using Eq.(\ref{wavegamma}) and property of two-body scattering amplitude, we find
\begin{align}
& T_{(\gamma)} (k ; q_{ij}, q_{k})   = -   \left [1 +  i t_{+}(\sqrt{\sigma^{2}     -\frac{3}{4} k^{2} } )  \right ]  \nonumber \\
& \quad \times  \int_{-\infty}^{\infty}  d r'_{\alpha \beta} d r'_{\gamma}    e^{ - i k r'_{\gamma} }   m V_{0}\delta (r'_{\alpha \beta}) \nonumber \\
& \quad       \times \left [ \psi_{(0)}(r'_{\alpha \beta}, r'_{\gamma};q_{ij}, q_{k})   + \psi_{(\alpha)}  (r'_{\beta\gamma}, r'_{\alpha};q_{ij}, q_{k}) \right.  \nonumber \\
& \quad \quad  \quad \quad \quad \quad    \quad  \quad    \quad       \quad    \quad     \left.  +  \psi_{(\beta)} (r'_{\gamma \alpha}, r'_{\beta};q_{ij}, q_{k})  \right ] , \nonumber \\
&\quad\quad \quad\quad \quad\quad \quad\quad  \quad\quad \quad\quad \quad     \alpha \neq \beta \neq \gamma , \label{Tgammawave}
\end{align}
therefore, the wave function $ \psi_{(\gamma)} $ is related to $ T_{(\gamma)} $ amplitude by
\begin{align}
 & \psi_{(\gamma)}   (r_{\alpha \beta}, r_{\gamma};q_{ij}, q_{k})   \nonumber   \\
&  \quad    =  i \int_{-\infty}^{\infty}   \frac{d q}{2\pi}    \frac{ e^{i \sqrt{\sigma^{2}    -\frac{3}{4} q^{2}} | r_{\alpha \beta} | }e^{i q  r_{\gamma}  }  }{2 \sqrt{\sigma^{2}    -\frac{3}{4} q^{2}}}   T_{(\gamma)}  ( q;q_{ij}, q_{k})       . \label{waveisobar}
\end{align}
Let's also define  functions $ v_{(\gamma)}$,
\begin{align}
& v_{(\gamma)} ( k ;q_{ij}, q_{k})  = \int_{-\infty}^{\infty}  d r_{\alpha \beta} d r_{\gamma}  e^{-i k  r_{\gamma}}   \nonumber \\
& \quad \quad \quad \quad   \quad \quad  \times m V_{0} \delta (r_{\alpha \beta})   \psi_{(0)} (r_{\alpha \beta}, r_{\gamma};q_{ij}, q_{k})  . \label{g0def}
\end{align}
   Eqs.(\ref{Tgammawave}-\ref{g0def}) all together thus yield coupled sets of integral equation of $T_{(\gamma)}$'s, which is exactly just Faddeev's equation for $\delta$-function potential,
\begin{align}
&T_{(\gamma)} (k ; q_{ij}, q_{k})   =  (2\sqrt{\sigma^{2}    -\frac{3}{4} k^{2} }) i t_{+}(\sqrt{\sigma^{2}    -\frac{3}{4} k^{2} } )    \nonumber \\
&  \quad   \times \left [ \frac{  v_{(\gamma)} (k; q_{ij}, q_{k}) }{ i m V_{0}}   \right. \nonumber \\
& \quad\quad \quad\quad \left.    + \quad   i \int_{-\infty}^{\infty}   \frac{d q }{2\pi}     \frac{  T_{(\alpha)}  (q; q_{ij}, q_{k})  + T_{(\beta)} (q; q_{ij}, q_{k})    }{ \sigma^{2}     -\frac{3}{4} q^{2} - (k  + \frac{q}{2})^{2}  + i \epsilon}     \right ], \nonumber \\
&\quad\quad \quad\quad \quad\quad \quad\quad  \quad\quad \quad\quad \quad  \quad\quad   \alpha \neq \beta \neq \gamma . \label{faddeevTeq}
\end{align}

At last, the total three-body scattering amplitude is defined by
\begin{align}
T  & (k_{\alpha \beta}, k_{\gamma} ; q_{ij}, q_{k}) = - \int_{-\infty}^{\infty}  d r_{\alpha \beta} d r_{\gamma}  e^{ - i k_{\alpha \beta} r_{\alpha \beta}} e^{-i k_{\gamma} r_{\gamma}}  \nonumber \\
& \quad   \times mV_{0}  \left [  \delta (r_{\alpha \beta})+\delta (r_{\beta \gamma})+\delta (r_{\gamma \alpha})  \right ] \psi( r_{\alpha \beta}, r_{\gamma} ; q_{ij}, q_{k})  . \label{total3bT}
\end{align}
As suggested in \cite{Faddeev:1960su,Faddeev:1965},  $T$ is thus represented as the sum of three $T_{(\gamma)}$ amplitudes,
 \begin{align}
T  (k_{\alpha \beta}, k_{\gamma}; q_{ij}, q_{k}) =  \sum_{\delta =1}^{3}T_{(\delta)} (k_{\delta}; q_{ij}, q_{k})  ,  \label{3bTsum}
\end{align}
where \mbox{$k_{\alpha} =- k_{\alpha \beta} -\frac{k_{\gamma}}{2}$} and \mbox{$k_{\beta} = k_{\alpha \beta} -\frac{k_{\gamma}}{2}$}.

As already mentioned in introduction,   unlike  two-body scattering, the solution of Faddeev's equation defined in Eq.(\ref{total3bT}) are not equivalent to the physical transition amplitudes  \cite{Faddeev:1965,Gerjuoy:1971,Osborn:1970zz,Osborn:1973zz}. The physical transition amplitudes for different physical processes  are associated to the residue functions of singularities of $T$-amplitude given in Eq.(\ref{total3bT}).  In general,  the   $T$-amplitude   has two distinct type singularities \cite{Faddeev:1965,Osborn:1970zz,Osborn:1973zz}. One type is called primary singularities, {\it e.g.} $\left (\sigma^{2} +\chi_{12}^{2} -\frac{3}{4} q^{2}_{3} \right)^{-1}$ where $\chi_{12}$ is  bound state energy of pair $(12)$. The pole of $\left (\sigma^{2} +\chi_{12}^{2} -\frac{3}{4} q^{2}_{3} \right)^{-1}$ type  presents in driving term of Faddev's equation and persists in all terms of an iterative series of amplitudes. It arise when relative momentum of $(12)$ pair hits   the bound state pole position of two-body scattering amplitude,  
\mbox{$t(q_{12}) \sim (q_{12}^{2}+\chi_{12}^{2})^{-1}$}, and it describes the possibility of existence  of two-body bound state in both initial and final states. The other types of singularity, called secondary singularities \cite{Faddeev:1965,Osborn:1970zz,Osborn:1973zz},  only present in driving terms and first a few iterations, and  singularities are getting weaker and eventually disappear after a couple of iterations.  A typical example is the $\delta$-functions that are related to  disconnected diagrams with   third particle remaining intact. The existence of   singularities in $T$-amplitude is the consequence of presence of multiple possible physical distinct processes in a three-body system. Hence, these singularities are directly  associated with the different physically realizable asymptotic states of the system. The explicit decomposition of primary singularities of   $T$-amplitude is given in \cite{Faddeev:1965,Osborn:1970zz,Osborn:1973zz} by
\begin{align}
&T   (k_{12}, k_{3} ; q_{12}, q_{3})   = \sum_{k=1}^{3}  (2\pi) \delta( k_{k} -q_{k}) (2 q_{ij}) t(k_{ij}; q_{ij}) \nonumber \\
&+ \sum_{\gamma, k=1}^{3} \left [   \mathcal{F}_{(\gamma,  k)} (k_{\alpha \beta}, k_{\gamma} ; q_{ ij }, q_{k})  \right. \nonumber \\
& \quad \quad\quad    \left.  + \frac{\phi_{(\gamma) } (k_{\alpha\beta}) \mathcal{G}^{*}_{(\gamma, k)} (k_{\gamma}; q_{ ij}, q_{k}) }{\sigma^{2}+ \chi_{\alpha \beta}^{2} - \frac{3}{4} k_{\gamma}^{2}}  \right. \nonumber \\
& \quad \quad\quad    + \frac{ \mathcal{G}_{(\gamma, k)} (k_{\alpha \beta}, k_{\gamma};  q_{k})  \phi^{*}_{(k) } (q_{ij})}{\sigma^{2}+ \chi_{ij}^{2} - \frac{3}{4} q_{k}^{2}} \nonumber \\
&\quad \quad\quad    \left. +  \frac{\phi_{(\gamma) } (k_{\alpha\beta}) \mathcal{K}_{(\gamma, k)} (k_{\gamma}; q_{k})  \phi^{*}_{(k) } (q_{ij})}{ \left( \sigma^{2}+ \chi_{\alpha \beta}^{2} - \frac{3}{4} k_{\gamma}^{2} \right) \left ( \sigma^{2}+ \chi_{ij}^{2} - \frac{3}{4} q_{k}^{2}\right )}  \right ], \label{singT}
\end{align}
where $t$ function in    Eq.(\ref{singT}) stands for two-body off-shell scattering amplitude, $\phi_{(\gamma)}$ function represents the vertex function of the two-body bound state wave function, \mbox{$\phi_{(\gamma)}(k_{\alpha \beta}) = (k_{\alpha\beta}^{2} + \chi_{\alpha \beta}^{2}) \psi (k_{\alpha \beta})$}. The residue functions that are associated to physical transition amplitudes, $\mathcal{F}_{(\gamma,  k)}$, $\mathcal{G}_{(\gamma,  k)}$ and $\mathcal{K}_{(\gamma,  k)}$  in Eq.(\ref{singT}), do not have any primary singularities, though they may still have secondary singularities. It has been shown in \cite{Faddeev:1965,Osborn:1970zz,Osborn:1973zz} that  the first term on the right hand side of Eq.(\ref{singT}) is from disconnected contribution and it describes the process that one of incident free particles is unscattered. The $\mathcal{K}_{(\gamma,  k)}$ function in last term  is the physical amplitude that describes  the processes of either direct or rearrangement     scattering on a bound state: \mbox{$(ij)+k\rightarrow (\alpha\beta)+\gamma $}. \mbox{$\sum_{\gamma=1}^{3} \left [ \mathcal{G}_{(\gamma, k)}+\frac{\phi_{(\gamma) }  \mathcal{K}_{\gamma, k}    }{   \sigma^{2}+ \chi_{\alpha \beta}^{2} - \frac{3}{4} k_{\gamma}^{2} }  \right ] $} is the physical transition amplitude for breakup or capture processes. The true \mbox{$1+2+3 \rightarrow 1+2+3$} physical scattering amplitude is given by on-shell $T$-amplitude in physical kinematic domain of three-particle momenta. The singularities in momentum space generate more complicated asymptotic form of three-body wave function than that of two-body wave function in configuration space. The physical transition amplitudes can thus also be   defined by asymptotic forms of wave function in configuration representation \cite{Nuttall:1969mp,Newton:1972xc,Osborn:1973zz,Merkurev:1976uy}. The asymptotic form of  three-body wave function 
depends on the type of initial state, and   may  behave quite differently and  describe distinct physical processes  at different domains in $(r_{ij}, r_{k})$ plane \cite{Nuttall:1969mp,Newton:1972xc,Merkurev:1976uy,Osborn:1970zz,Osborn:1973zz}. For a example,    scattering of 3rd particle on a bound state of $(12)$ pair,   the physical amplitudes of different processes are given by asymptotic wave function in different domains:  ($i$) direct channel scattering, \mbox{$(12)+3 \rightarrow (12) +3$}, is described   in domain of   finite $r_{12}$  and large $r_{3}$, the scattering part of  asymptotic wave function is of the oder of \mbox{$ e^{- \chi_{12} |r_{12}|} \mathcal{O} (|r_{3}|^{-1})$}; ($ii$) rearrangement scattering processes,  \mbox{$(12)+3 \rightarrow (23)+1$} or \mbox{$(12)+3 \rightarrow (31)+2$}, are given   in domains of finite $r_{23}$ and large $r_{1}$ or finite $r_{31}$ and large $r_{2}$ respectively, and the wave function behave as \mbox{$ e^{- \chi_{23} |r_{23}|} \mathcal{O} (|r_{1}|^{-1})$} or \mbox{$ e^{- \chi_{31} |r_{31}|} \mathcal{O} (|r_{2}|^{-1})$} respectively; ($iii$) breakup process, \mbox{$(12)+3 \rightarrow 1+2+3$},  appears  as both   $r_{12}$ and $r_{3}$ are large, but \mbox{$r_{12}/r_{3}$} remains constant \cite{Osborn:1970zz,Osborn:1973zz}, the wave function for breakup is of the order of \mbox{$  \mathcal{O} \left ((r_{12}^{2}+ \frac{4}{3} r_{3}^{2})^{-\frac{5}{4}} \right)$}.    In the case of three free incident particles \cite{Nuttall:1969mp,Newton:1972xc,Merkurev:1976uy},  the asymptotic form of wave function consists of several different pieces that decrease at different rates and describes different distinct physical processes, and its fall-off also depends on the direction in configuration space:  (1) incident plane wave that does not decrease in any direction; (2) terms describe the scattering of a pair by itself without participation of third particle,  the wave function is of the order of $\mathcal{O}(|r_{ij}|^{-1})$  (k-th particle as spectator). The disconnected terms and incident free wave must be subtracted out before other contributions become visible; (3) the terms that describe capture of pair $(ij)$ as bound state is of the order of \mbox{$ e^{- \chi_{ij} |r_{ij}|} \mathcal{O} (|r_{k}|^{-1})$}. (4) the terms  generated by on-shell double scattering is of the order of $\mathcal{O} \left (  (r_{12}^{2}+ \frac{4}{3} r_{3}^{2})^{-1} \right )$; (5) The true three-body scattering terms are of the order of  \mbox{$  \mathcal{O} \left ((r_{12}^{2}+ \frac{4}{3} r_{3}^{2})^{-\frac{5}{4}} \right)$}. Nevertheless, it is clear that understanding of either singularities structure of $T$-amplitude in momentum space or asymptotic form of wave function in configuration space is crucial step in order to extract physical transition amplitudes.

As mentioned early in introduction, McGuire's model permits  no diffraction and no breakup or capture processes, only forward scattering (no new momenta are created after collision). Therefore, the physical processes in McGuire's model are split into two decoupled sectors: scattering on a bound state and three-to-three particles scattering. The analytic solution of Faddeev's equation for a particle scattering on a bound state has been discussed in \cite{Dodd:1970,Majumdar:1972,Gerjuoy:1985}. 
In this work,  we solve Faddeev's equation Eq.(\ref{faddeevTeq})   analytically for \mbox{$1+2+3 \rightarrow 1+2+3$}  three   particles scattering process.  As will be shown in following sections, the  $T$-amplitude for three-to-three  scattering of identical bosons   in McGuire's model has the form of
\begin{align}
 & T   (k_{12}, k_{3} ;  q_{12}, q_{3})  \nonumber \\
 & =    \sum_{\gamma=1}^{3} 2 (2\pi) \delta( k_{3} -q_{\gamma})    \sum_{k=1}^{3} (2  \sqrt{\sigma^{2} -\frac{3}{4} q_{k}^{2}} )t_{+}(  \sqrt{\sigma^{2} -\frac{3}{4} q_{k}^{2}} )  \nonumber \\
&   + \sum_{\gamma=1}^{3}  \frac{R(k_{\gamma})}{(k_{\gamma}-q_{1})(k_{\gamma}-q_{2})(k_{\gamma}-q_{3})}  , \label{mcguireT}
\end{align}
where the first term again represent the disconnected contribution and second term represents the sum of all the rescattering effect, and $R(k)$ function is free of poles. The pole structure in Eq.(\ref{mcguireT}) yield the forward scattering of three particles in the end. Notice that the exact solution of Faddeev's equation in Eq.(\ref{mcguireT}) has only singularities of poles and $\delta$-function. This means that first of all, the branch cut contribution during the iteration of Faddeev's equation has to be all cancelled out, it turns out to be true, see Section \ref{Tsolindv}. Secondly, higher order iterations of Faddeev's equation in one dimension do not completely eliminate the   three-particle propagator   singularities, this is distinct feature from three dimensional three-body physics. In both dimensions, off-shell  double scattering   display the similar singularity structure of three-particle propagator, {\it e.g.} $\left [ q_{12}^{2}-(k_{3} + \frac{q_{3}}{2} )^{2}+ i \epsilon  \right ]^{-1}$,  see Fig.\ref{fig:feynman}(b) and Fig.\ref{fig:feynman}(c). However, for triple scattering, the singularity structure starts diverging. A triple scattering in three dimension has the typical singularities of type,
\begin{align}
&  \int   d^{3} q \frac{1}{\sigma^{2} -\frac{3}{4} \mathbf{ k}_{3}^{2} -(\mathbf{ q} + \frac{\mathbf{ k}}{2})^{2} + i \epsilon}    \frac{1}{\mathbf{ q}_{12}^{2}-(\mathbf{ k}_{3} + \frac{\mathbf{ q}_{3}}{2} )^{2}+ i \epsilon } \nonumber \\
& = \frac{i \pi^{2}}{ |\frac{\mathbf{ k}_{3}}{2} -\frac{\mathbf{ q}_{3}}{2} |} \ln \frac{\mathbf{ q}^{2}_{12} + \sqrt{\sigma^{2} -\frac{3}{4} \mathbf{ k}_{3}^{2}} +|\frac{\mathbf{ k}_{3}}{2} -\frac{\mathbf{ q}_{3}}{2} |^{2}  }{\mathbf{ q}^{2}_{12} + \sqrt{\sigma^{2} -\frac{3}{4} \mathbf{ k}_{3}^{2}} -|\frac{\mathbf{ k}_{3}}{2} -\frac{\mathbf{ q}_{3}}{2} |^{2} }.
\end{align}
In one dimension, triple scattering appears as a one dimensional integral over the product of two three-particle propagators,
\begin{align}
&   \int_{-\infty}^{\infty}   d q    \frac{  1  }{ \sigma^{2}   -\frac{3}{4}  k_{3}^{2} - (q + \frac{k_{3}}{2})^{2}  + i \epsilon}      \frac{  1  }{  q_{12}^{2} - (q  + \frac{q_{3}}{2})^{2} + i \epsilon }    ,
\end{align}
it is easy to see that after picking up the poles in integrand, the result of one dimensional integral  has only poles and branch cuts. As shown in \cite{Faddeev:1965}, in three dimension,  three-particle propagator singularities are   smoothed out in higher iterations  and are thus considered as secondary singularities. On the contrary, in one dimension, some poles survived higher iterations and all the branch cuts are cancelled out after sum over all the diagrams in McGuire's model. In the end, exact solution of Faddeev's equation displays only singularities of poles and $\delta$-function, as in Eq.(\ref{mcguireT}).

On the other hand,   it will also be   shown in Section \ref{3bosonsfree} that the principal part of pole term in Eq.(\ref{mcguireT}) is proportional to a factor, \mbox{$(\sigma^{2} - k^{2}_{12} - \frac{3}{4} k^{2}_{3})$}. As the consequence, the solutions of Faddeev's equation suffer no branch cut singularities, all the branch cut singularities in Faddeev's equation are  cancelled out.   The three-body wave function consists of only six plane waves: $e^{i q_{ij} r_{12}} e^{i q_{k} r_{3}}$ $(k=1,2,3)$,  no diffraction effect is generated after scattering.  When three-particle $T$-amplitude is put on energy shell: \mbox{$\sigma^{2} = k^{2}_{12} + \frac{3}{4} k^{2}_{3}$},  the principal part of pole term vanishes, thus,  the on-shell physical three-body amplitude consist of only the  terms that is proportional to $\delta$-function from both disconnected diagrams and on-shell three-body rescattering effect. Therefore, it allows us to define the  on-shell scattering physical amplitude  as residue of $T$-amplitude at pole positions,
\begin{align}
 &\left( - \sum_{\gamma=1}^{3} 2 (2\pi) \delta(  k-q_{\gamma}) \right)  (2 \sqrt{\sigma^{2} - \frac{3}{4}k^{2}})\mathcal{T} (k )   \nonumber \\
& \quad \quad\quad    =    T    (\sqrt{\sigma^{2} - \frac{3}{4}k^{2}}, k  ;   q_{12}, q_{3})   , \ \ \ \    k = q_{1,2,3}. \label{forwardT}
\end{align}

  In general, there are six possible independent incoming plane waves in terms of  permutation of incoming momenta, see  \cite{McGuire:1964zt}. In Appendix \ref{TsolFaddeev}, we show details of how Faddeev's equation is solved for a incoming plane wave \mbox{$\psi_{(0)}=e^{i q_{12} r_{12}}e^{i q_{3} r_{3}}$} as an example. In the end of Appendix \ref{Tsolindv}, the analytic solutions of Faddeev equation  for scattering amplitudes $T_{(\gamma)}$'s and physical on-shell $S$-matrix are presented for all six possible incoming plane waves. The three-body wave function is constructed by using solution of Faddeev's equation, $T_{(\gamma)}$'s,   we also present the result of constructed three-body wave function for incoming plane wave \mbox{$\psi_{(0)}=e^{i q_{12} r_{12}}e^{i q_{3} r_{3}}$} as an example in Appendix \ref{constwave}.  Although,  there are six independent sets of solutions of Faddeev's equation corresponding to six independent incoming plane waves,  for three  spinless identical particles, only solutions for totally symmetric and totally anti-symmetric wave functions have meaningful physical correspondences: scattering of three spinless bosons and three spinless fermions respectively. Hence, in follow sections,  the attentions are focused on three spinless bosons and three spinless fermions scattering only.

\subsection{Solutions of Faddeev's equation for three spinless   fermions}\label{3fermionsfree}
For three spinless identical fermions, the wave function has to be totally anti-symmetric under exchange of any two particles coordinates,   the free incoming wave for totally anti-symmetric three fermions   is 
\begin{equation}
\psi^{anti}_{(0)} =\sum_{k=1}^{3} \left ( e^{i q_{ij} r_{12}}-e^{-i q_{ij} r_{12}} \right )  e^{ i q_{k} r_{3}}.\label{psi0sym}
\end{equation}
 Given the solutions of scattering amplitudes for each individual wave in Section \ref{Tsolindv}, it is easy to see that   the solutions of  Faddeev's equation for three spinless identical fermions  all vanish, \mbox{$ T_{(\gamma)}  = 0, \gamma=1,2,3$}. Therefore, the totally anti-symmetric wave function for three identical fermions is given by free incoming wave solution, \mbox{$ \psi_{anti} (r_{12},r_{3}; q_{ij}, q_{k})=\psi^{anti}_{(0)} $}.

\subsection{Solutions of Faddeev's equation for three spinless   bosons}\label{3bosonsfree}
For three spinless identical bosons, the three-body wave function has to be totally symmetric under exchange of arbitrary two particles coordinates, the free incoming wave for totally symmetric three bosons  is given by
\begin{equation}
\psi^{sym}_{(0)} =\sum_{k=1}^{3} \left ( e^{i q_{ij} r_{12}}+e^{-i q_{ij} r_{12}} \right )  e^{ i q_{k} r_{3}}, \label{psi0sym}
\end{equation}
therefore,  
\begin{align}
v_{(1,2,3)} (k; q_{12}, q_{3})= m V_{0} \sum_{k=1}^{3} 2 (2\pi)    \delta(k-q_{k}) .
\end{align}
Again using the solutions of scattering amplitudes for each individual wave in Section \ref{Tsolindv}, we found that the solution of  Faddeev's equation for three identical bosons  are 
  \begin{align}
 T_{(1,2,3)}  &( k; q_{ij} ,q_{k} ) =  2 (2 \pi i ) \delta(k -q_{1})   \frac{   i m V_{0}   }{ 1- \frac{ i  m V_{0}}{2 q_{23}}     }  \nonumber \\
 & + 2 (2 \pi i ) \delta(k-q_{2})   \frac{  i m V_{0}   }{  1- \frac{ i  m V_{0}}{2 q_{31}}  } \nonumber \\
 & + 2 (2 \pi i ) \delta(k-q_{3})  \frac{    i m V_{0} \left (1+ \frac{ i  m V_{0}}{2 q_{23}} \right )     }{\left (1+ \frac{ i  m V_{0}}{2 q_{12}} \right ) \left (1- \frac{ i  m V_{0}}{2 q_{23}} \right ) }  \nonumber \\
 &      -   2 \frac{   \frac{   6 \left (  i m V_{0}   \right )^{2}    k  }{\left (1+ \frac{ i  m V_{0}}{2 q_{12}} \right ) \left (1- \frac{ i  m V_{0}}{2 q_{23}} \right ) \left (1- \frac{ i  m V_{0}}{2 q_{31}} \right ) } }{ \left ( k- q_{3} -i \epsilon  \right ) \left (k - q_{2} - i \epsilon  \right )  \left (k- q_{1} + i \epsilon  \right )}                  .   \label{T123sym}
 \end{align} 
All three $T_{(\gamma)}$'s are identical due to Bose symmetry.   By picking up the contribution of poles, \mbox{$k=q_{2}+ i \epsilon$}, \mbox{$q_{3}+ i \epsilon$}, and \mbox{$q_{1}- i \epsilon$} in Eq.(\ref{T123sym}), we introduce three on-shell   scattering amplitudes for later convenience of presentation,
 \begin{align}
 i \mathcal{T}_{3} &=\frac{    \left ( - \frac{i m V_{0} }{2 q_{12}}\right ) \left (1- \frac{ i  m V_{0}}{2 q_{23}} \frac{ i  m V_{0}}{2 q_{31}} \right )     }{  \left ( 1+ \frac{ i  m V_{0}}{2 q_{12}} \right ) \left (1- \frac{ i  m V_{0}}{2 q_{23}} \right ) \left (1- \frac{ i  m V_{0}}{2 q_{31}} \right )  }    , \nonumber \\
  i  \mathcal{T}_{1}    &=\frac{  \left ( \frac{i m V_{0}}{2 q_{23}} \right )    \left (1- \frac{ i  m V_{0}}{2 q_{31}}  \frac{ i  m V_{0}}{2 q_{12}} \right ) }{  \left ( 1+ \frac{ i  m V_{0}}{2 q_{12}} \right ) \left (1- \frac{ i  m V_{0}}{2 q_{23}} \right ) \left (1- \frac{ i  m V_{0}}{2 q_{31}} \right )  }     ,  \nonumber \\
  i \mathcal{T}_{2}    &=\frac{  \left ( \frac{i m V_{0}}{2 q_{31}} \right ) \left (1- \frac{ i  m V_{0}}{2 q_{23}}  \frac{ i  m V_{0}}{2 q_{12}} \right )  }{\left (1+ \frac{i m V_{0}}{2 q_{12}} \right )\left (1- \frac{i m V_{0}}{2 q_{23}} \right )\left (1- \frac{i m V_{0}}{2 q_{31}} \right )}    , \label{reducedTgamma}
 \end{align}
 and
 \begin{align}
 i \mathbf{ T}_{1} & =  i  \mathcal{T}_{1} -  i t_{+} (-q_{23}), \ \ 
  i \mathbf{ T}_{2}  = i  \mathcal{T}_{2} - i t_{+} (-q_{31}), \nonumber \\
 i \mathbf{ T}_{3}&=  i  \mathcal{T}_{3}  - i t_{+} (q_{12}) \left (1 + 2 i t_{+}(-q_{23})   \right ) = i \mathbf{ T}_{2} -  i \mathbf{ T}_{1}  , 
 \end{align}
where $t_{+}$ again is two-body scattering amplitude given in Eq.(\ref{deltatamp}).
The on-shell   scattering contribution of $T_{(\gamma)}$  is thus given by
\begin{align}
&  T^{(phys)}_{(\gamma)}  ( k;   q_{ij} ,q_{k} )  =  2 (2 \pi i ) \delta(k -q_{1}) (2 q_{23}) i   \mathcal{T}_{1} \nonumber \\
&    + 2 (2 \pi i ) \delta(k -q_{2}) (2 q_{31}) i  \mathcal{T}_{2} - 2 (2 \pi i ) \delta(k -q_{3}) (2 q_{12}) i   \mathcal{T}_{3} .
\end{align}
The on-shell     amplitudes $\mathcal{T}_{\gamma}$ in Eq.(\ref{reducedTgamma}) satisfy unitarity relations,
\begin{align}
\mbox{Im} \mathcal{T}_{\gamma} = \mathcal{T}_{\gamma}^{*} (\mathcal{T}_{3} + \mathcal{T}_{1}+\mathcal{T}_{2}), \ \ \gamma =1,2,3. \label{3bunit}
\end{align}
The three-body off-shell scattering amplitude thus reads,
\begin{align}
T(k_{12}, & k_{3}; q_{ij}, q_{k}) = \sum_{\gamma=1}^{3}   T^{(phys)}_{(\gamma)}  ( k_{\gamma};   q_{ij} ,q_{k} )  \nonumber \\
&- 2\sum_{\gamma=1}^{3}     \mathcal{P} \frac{   \frac{   6 \left (  i m V_{0}   \right )^{2}    k_{\gamma}  }{\left (1+ \frac{ i  m V_{0}}{2 q_{12}} \right ) \left (1- \frac{ i  m V_{0}}{2 q_{23}} \right ) \left (1- \frac{ i  m V_{0}}{2 q_{31}} \right ) } }{ \left ( k_{\gamma}- q_{3}  \right ) \left (k_{\gamma} - q_{2}   \right )  \left (k_{\gamma}- q_{1}   \right )}, \label{offshellT}
\end{align}
where $\mathcal{P}$ stands for the principal part of poles, and \mbox{$k_{1} = - k_{12} - \frac{1}{2} k_{3}$} and \mbox{$k_{2} =  k_{12} - \frac{1}{2} k_{3}$}.  It is easy to show that
\begin{align}
\sum_{\gamma=1}^{3}     \mathcal{P} \frac{       k_{\gamma}    }{ \left ( k_{\gamma}- q_{3}  \right ) \left (k_{\gamma} - q_{2}   \right )  \left (k_{\gamma}- q_{1}   \right )} \propto (\sigma^{2} - k_{12}^{2} - \frac{3}{4} k_{3}^{2}), 
\end{align}
therefore, principal  part on right hand side  of Eq.(\ref{offshellT}) vanishes for on-shell scattering amplitude.
The Bose symmetry warrants that all six on-shell $S$-matrix are identical and given by \mbox{$S_{sym}  = 1+ 2     i \mathcal{T}   $}, where  as defined in Eq.(\ref{forwardT}),  \mbox{$\mathcal{T} $} is physical   scattering amplitude, and  \mbox{$\mathcal{T}= \sum_{k=1}^{3}   \mathcal{T}_{k}$}, thus, we find
\begin{align}
S_{sym}     =   \frac{   \left (1- \frac{ i  m V_{0}}{2 q_{12}} \right ) \left (1+ \frac{ i  m V_{0}}{2 q_{23}} \right ) \left (1+ \frac{ i  m V_{0}}{2 q_{31}} \right )  }{\left (1+ \frac{ i  m V_{0}}{2 q_{12}} \right ) \left (1- \frac{ i  m V_{0}}{2 q_{23}} \right ) \left (1- \frac{ i  m V_{0}}{2 q_{31}} \right ) }  .
\end{align}

The on-shell physical   scattering amplitudes and $S$-matrix can be expressed in terms of a single   two-body scattering phase shift, \mbox{$ t_{+}(q) = \frac{e^{2 i \delta (q) }-1}{2i }$},  thus, we obtain
  \begin{align}
 i \mathcal{T}_{3} &= \left ( \frac{e^{2 i \delta (q_{12}) }-1}{2}  \right ) \left ( \frac{e^{ 2 i \delta (-q_{23})} + e^{ 2 i \delta (-q_{31})} }{2}  \right ),     \nonumber \\
  i  \mathcal{T}_{1}   &  = \left ( \frac{e^{2 i \delta (-q_{23})}-1}{2}  \right ) \left ( \frac{1+e^{ 2 i \delta (-q_{31}) }  e^{ 2 i \delta (q_{12})} }{2}  \right )  ,  \nonumber \\
  i \mathcal{T}_{2}    &= \left ( \frac{e^{2 i \delta (-q_{31})}-1}{2}  \right ) \left ( \frac{1+e^{ 2 i \delta (-q_{23})}  e^{ 2 i \delta (q_{12})} }{2}   \right )  ,  \nonumber \\
  S_{sym} &=e^{2 i \left ( \delta (q_{12}) + \delta(-q_{23}) + \delta(-q_{31})   \right )}.  \label{phaseshiftT}
 \end{align}
 It can be easily checked that the phase shift expressions of  on-shell   amplitudes $\mathcal{T}_{\gamma}$ in Eq.(\ref{phaseshiftT}) are the consequence of unitarity relations in Eq.(\ref{3bunit}), therefore, the Eq.(\ref{phaseshiftT}) may be more general for pair-wise and short range interactions of three identical particles scattering.

 The totally symmetric wave function can be constructed by using Eq.(\ref{waveisobar}) and solutions of Faddeev's equation given in Eq.(\ref{T123sym}). An example of construction of  wave function from solutions of Faddeev's equation is given in Appendix \ref{constwave}, the construction is rather length and tedious, so we do not present all the details in text.  The totally symmetric wave function is expressed in terms of a single independent coefficient,
 \begin{align}
& \psi_{sym} (r_{12},r_{3}; q_{ij}, q_{k}) - \psi^{sym}_{(0)}  \nonumber \\
& =\left ( A_{sym}   (r_{12},r_{3}) e^{i q_{12} r_{12}} + A_{sym}   (-r_{12},r_{3})  e^{ - i q_{12} r_{12}} \right ) e^{i q_{3} r_{3}}  \nonumber \\
&+ \left ( A_{sym}   (r_{31},r_{2}) e^{i q_{23} r_{12}} + A_{sym}  (-r_{23},r_{1}) e^{ - i q_{23} r_{12}} \right ) e^{i q_{1} r_{3}} \nonumber \\
&+ \left ( A_{sym}   (r_{23},r_{1})  e^{i q_{31} r_{12}} + A_{sym}  (-r_{31},r_{2}) e^{ - i q_{31} r_{12}} \right ) e^{i q_{2} r_{3}}, \label{wavesolsym}
\end{align}
where \mbox{$r_{23} =  -\frac{r_{12}}{2} + r_{3} $}  and  \mbox{$r_{31} =  -\frac{r_{12}}{2} - r_{3}$}, and
\begin{align}
A_{sym}  & (r_{12},r_{3}) = 1 + \theta(r_{12})   2  it_{+}(q_{12}) \left [ 1+ 2 i t_{+}(-q_{23})\right ]   \nonumber \\
&+    \theta(-r_{23})  2  it_{+}(-q_{23}) +  \theta(-r_{31})  2  it_{+}(-q_{31})      \nonumber \\
  &-     \theta(r_{12})  \theta(r_{23}) 4    i  \mathbf{ T}_{1}      + \theta(r_{12}) \theta(-r_{31})   4   i  \mathbf{ T}_{2}         . \label{coefwavesolsym}
\end{align}

\section{Three-body scattering in finite volume}\label{3bbox}
For three particles interaction in a one dimensional periodic box with the size of $L$,   the wave function of three-particle in finite volume, \mbox{$\Psi^{(L)}(x_{1},x_{2},x_{3}; p_{1},p_{2},p_{3})$},   must   satisfy periodic boundary condition,
\begin{align}
& \Psi^{(L)}(x_{1} + n_{x_{1}} L,x_{2} + n_{x_{2}} L,x_{3}+ n_{x_{3}} L; p_{1},p_{2},p_{3})  \nonumber \\
& \quad \quad  \quad \quad  = \Psi^{(L)}(x_{1},x_{2},x_{3}; p_{1},p_{2},p_{3}), \ \ \ \ n_{x_{1}, x_{2},x_{3}} \in \mathbb{Z}. \label{3bperiodic}
\end{align} 
The finite volume three-body wave function, $\Psi^{(L)}$, is constructed from three-body free space wave function, $\Psi$, by
\begin{align}
& \Psi^{(L)}(x_{1},x_{2},x_{3}; p_{1},p_{2},p_{3})  =  \frac{1}{V} \sum_{n_{x_{1}}, n_{ x_{2} }, n_{x_{3}}  \in \mathbb{Z}}   \nonumber \\
 & \quad \quad \times \Psi(x_{1} + n_{x_{1}} L,x_{2} + n_{x_{2}} L,x_{3}+ n_{x_{3}} L; p_{1},p_{2},p_{3})    ,
\end{align}
in this way, the periodic boundary condition in Eq.(\ref{3bperiodic}) is warranted.
Factorizing center of mass wave function and relative wave function, and also defining new variables, \mbox{$n = n_{x_{k}}$}, \mbox{$n_{ij} = n_{x_{i}}-n_{x_{j}}$} and \mbox{$n_{k} = (n_{x_{i}}+n_{x_{j}}) -2 n_{x_{k}}$}, where \mbox{$(n, n_{ij}, n_{k}) \in \mathbb{Z}$},  thus, we find
\begin{align}
& \Psi^{(L)}(x_{1},x_{2},x_{3}; p_{1},p_{2},p_{3})   \nonumber \\
& \quad \quad \quad \quad  \quad   = \frac{1}{V} \left ( \sum_{n  \in \mathbb{Z}} e^{ i P n  L} \right )  e^{ i P R}  \psi^{(L)} (r_{ij}, r_{k}; q_{ij}, q_{k}), \nonumber \\
&  \psi^{(L)} (r_{ij}, r_{k}; q_{ij}, q_{k})  \nonumber \\
& \quad \quad =  \sum_{n_{ij}, n_{k} \in \mathbb{Z}} e^{i \frac{P}{3} n_{k} L } \psi(r_{ij} + n_{ij} L, r_{k} + \frac{1}{2} n_{k} L; q_{ij}, q_{k}),  \label{relwaveconstr}
\end{align}
where $\psi^{(L)}   $  represents the relative finite volume wave function, and the normalization factor of infinite summation $V$ is given by, \mbox{$V= \sum_{n  \in \mathbb{Z}} e^{ i P n  L} = \frac{2\pi}{L}  \sum_{d \in \mathbb{Z}} \delta (P+\frac{2\pi}{L} d)$}.  The integer variables \mbox{$(n_{ij},n_{k})$} are related to relative coordinates \mbox{$(r_{jk},r_{k})$}  in $(ij)k$ configuration. The   relative variables  in other configurations can be expressed in terms of variables \mbox{$(n_{ij},n_{k})$}, {\it e.g.}  integer variables \mbox{$(n_{jk},n_{i})$} in $(jk)i$ configuration are given by,
\begin{align}
n_{jk} = - \frac{1}{2} n_{ij} + \frac{1}{2} n_{k}, \ \  n_{i} = - \frac{3}{2} n_{ij} - \frac{1}{2} n_{k}, \ \  i \neq j \neq k.
\end{align}
After removal of center of mass motion, the periodic boundary condition for relative finite volume wave function now reads,
\begin{align}
& \psi^{(L)}(r_{ij} + n_{ij} L,r_{k} +\frac{1}{2} n_{k} L ; q_{ij}, q_{k})  \nonumber \\
& \quad \quad   =  e^{ - i \frac{P}{3} n_{k} L} \psi^{(L)}(r_{ij}  ,r_{k} ; q_{ij}, q_{k} ) , \ \ P= \frac{2\pi}{L} d , d \in \mathbb{Z}  . \label{finitewavefunc}
\end{align}
With the solution of wave functions, for instance,  the  totally symmetric wave function given in Eq.(\ref{wavesolsym}-\ref{coefwavesolsym}),  the finite volume  three-body  wave function is constructed by using Eq.(\ref{relwaveconstr}). The infinite summation   in one dimension can be  performed by using the property of geometric series 
\begin{align}
 \sum_{n=\alpha}^{\infty} x^{n} = \frac{x^{\alpha}}{1-x}, \ \ \ \  (n, \alpha) \in \mathbb{Z}.
 \end{align}
   Hence the analytic solutions in one dimension for $\delta$-function potential can be obtained. As  have been mentioned in previous sections, not all six independent wave functions are   corresponding to physical systems, except    totally anti-symmetric  and symmetric wave functions given in Section \ref{3fermionsfree} and \ref{3bosonsfree}, which represent three identical fermions and bosons scattering respectively. The other four wave functions are not related to any physical processess, and indeed, we found no physical solutions in finite volume except three spinless bosons and fermions systems. In the case of three spinless identical fermions, because two-body interaction by $\delta$-function potential  vanishes for identical spinless fermions, three identical fermions experience  zero scattering effect, and behave as free particles. Therefore, the three-body wave function has trivial solution in   free space, as shown in Section \ref{3fermionsfree}.   In finite volume, periodic boundary condition leads to the quantization of the momenta of three fermions   as  free particle in a finite box, \mbox{$p_{i} = \frac{2\pi}{L} n_{x_{i}}, n_{x_{i}} \in \mathbb{Z}$}. Nevertheless, in follows,  we will only work out all the details of the  finite volume wave function for three identical  bosons system.

In the case of three spinless identical bosons, the  expression of three-body wave function is dramatically simplified by symmetry consideration, the three-body wave function in free space is expressed by a single independent coefficient only, see Eq.(\ref{wavesolsym}-\ref{coefwavesolsym}). Therefore we only need to perform the infinite sum for a single plane wave, the rest of components of finite volume wave function are easily obtained by symmetry consideration. For instance, we can pick the plane wave \mbox{$ e^{iq_{12} r_{12} } e^{ i q_{3} r_{3}}$} component, the corresponding coefficient of plane wave in finite volume is then given by
\begin{align}
 & A^{(L)}_{sym} (r_{12},r_{3}) = \sum_{n_{12}, n_{3} \in \mathbb{Z}}    e^{iq_{12} n_{12} L} e^{ i \left (\frac{P}{3} +  \frac{q_{3} }{2} \right ) n_{3} L }\nonumber \\
 & \quad \quad  \times A_{sym}  (r_{12} + n_{12} L, r_{3} + \frac{1}{2} n_{3} L ) .
\end{align}
For non-trivial solutions, only last two terms  in Eq.(\ref{coefwavesolsym}) survive in finite box.  

First of all, for the term proportional to \mbox{$\theta(r_{12})\theta(r_{23})$}  in Eq.(\ref{coefwavesolsym}), the infinite sum reads
\begin{align}
 &   \sum_{n_{12}, n_{3} \in \mathbb{Z}}^{n_{23} =  \frac{n_{3} - n_{12}}{2}} e^{iq_{12} n_{12} L} e^{ i \left (\frac{P}{3} +  \frac{q_{3} }{2} \right ) n_{3} L }    \theta(r_{12} + n_{12} L) \theta( r_{23} +  n_{23}  L )    \nonumber \\
 &= \sum_{ n_{12} = \theta(-r_{12})}^{\infty} e^{ -  i ( \frac{2}{3} P + q_{1}) n_{12} L  }  \sum_{ n_{23} = \theta(-r_{23}) }^{\infty} e^{  i ( \frac{2}{3} P + q_{3}) n_{23} L  } \nonumber \\
 & = \left [ \theta(r_{12})  + \frac{ e^{ -  i ( \frac{2}{3} P + q_{1})  L  } }{1- e^{ -  i ( \frac{2}{3} P + q_{1})   L  } } \right ] \left [ \theta(r_{23}) + \frac{ e^{  i ( \frac{2}{3} P + q_{3})  L  } }{1- e^{   i ( \frac{2}{3} P + q_{3})   L  } } \right].
\end{align}

Next, for the term proportional to  \mbox{$\theta(r_{12}) \theta(-r_{31})  $}  in Eq.(\ref{coefwavesolsym}), we have
\begin{align}
 &   \sum_{n_{12}, n_{3} \in \mathbb{Z}}^{n_{31} = - \frac{n_{3} +n_{12}}{2}} e^{iq_{12} n_{12} L} e^{ i \left (\frac{P}{3} +  \frac{q_{3} }{2} \right ) n_{3} L }   \nonumber \\
 & \quad \quad \quad \quad \quad \quad  \times  \theta(r_{12} + n_{12} L) \theta( -r_{31} - n_{31}  L )    \nonumber \\
 &= \sum_{ n_{12} = \theta(-r_{12})}^{\infty} e^{   i ( \frac{2}{3} P + q_{2}) n_{12} L  }  \sum_{ n_{31} =-\infty  }^{-\theta(r_{31})} e^{ - i ( \frac{2}{3} P + q_{3}) n_{31} L  } \nonumber \\
 & = \left [ \theta(r_{12})  + \frac{ e^{   i ( \frac{2}{3} P + q_{2})  L  } }{1- e^{   i ( \frac{2}{3} P + q_{2})   L  } } \right ] \left [ \theta(-r_{31}) + \frac{ e^{  i ( \frac{2}{3} P + q_{3})  L  } }{1- e^{   i ( \frac{2}{3} P + q_{3})   L  } } \right].
\end{align}

Putting everything together, we obtain finite volume coefficient of plane wave  \mbox{$ e^{iq_{12} r_{12} } e^{ i q_{3} r_{3}}$},
\begin{align}
& A^{(L)}_{sym}  (r_{12},r_{3})   = -     \theta(r_{12})  \theta(r_{23}) 4    i  \mathbf{ T}_{1}      + \theta(r_{12}) \theta(-r_{31})   4   i  \mathbf{ T}_{2}   \nonumber \\
&  -     \theta(r_{23})  4    i  \mathbf{ T}_{1}   \frac{ e^{- i (\frac{2}{3} P +q_{1}) L}}{1- e^{- i (\frac{2}{3} P +q_{1})}}   +  \theta(-r_{31}) 4   i  \mathbf{ T}_{2}     \frac{ e^{ i (\frac{2}{3} P +q_{2})L}}{1- e^{ i (\frac{2}{3} P +q_{2})L}}   \nonumber \\
&   + \theta(r_{12}) 4   i  \mathbf{ T}_{3}   \frac{  e^{ i (\frac{2}{3} P +q_{3}) L}}{1- e^{ i (\frac{2}{3} P +q_{3}) L}}    \nonumber \\
& + 4    i  \mathbf{ T}_{2}     \frac{ e^{ i (\frac{2}{3} P +q_{2})L}}{1- e^{ i (\frac{2}{3} P +q_{2})L}}         \frac{ e^{ i (\frac{2}{3} P +q_{3})L}}{1- e^{ i (\frac{2}{3} P +q_{3})L}}     \nonumber \\
&       -     4     i  \mathbf{ T}_{1}   \frac{ e^{- i (\frac{2}{3} P +q_{1})L}}{1- e^{- i (\frac{2}{3} P +q_{1})L}}        \frac{ e^{ i (\frac{2}{3} P +q_{3})L}}{1- e^{ i (\frac{2}{3} P +q_{3})L}}   . \label{boxcoefwavesolsym}
\end{align}
The coefficients for other waves in finite volume are obtained by symmetry consideration, {\it e.g.} for plane wave \mbox{$ e^{-iq_{12} r_{12} } e^{ i q_{3} r_{3}}$}, coefficient is given by \mbox{$A^{(L)}_{sym} (-r_{12},r_{3})  $}, {\it etc.} Therefore,    the three-body wave function for spinless bosons system  in finite volume yields 
 \begin{align}
& \psi^{(L)}_{sym} (r_{12},r_{3}; q_{ij}, q_{k})  \nonumber \\
& =\left ( A^{(L)}_{sym}   (r_{12},r_{3}) e^{i q_{12} r_{12}} + A^{(L)}_{sym}    (-r_{12},r_{3})  e^{ - i q_{12} r_{12}} \right ) e^{i q_{3} r_{3}}  \nonumber \\
&+ \left ( A^{(L)}_{sym}    (r_{31},r_{2}) e^{i q_{23} r_{12}} + A^{(L)}_{sym}  (-r_{23},r_{1}) e^{ - i q_{23} r_{12}} \right ) e^{i q_{1} r_{3}} \nonumber \\
&+ \left ( A^{(L)}_{sym}   (r_{23},r_{1})  e^{i q_{31} r_{12}} + A^{(L)}_{sym}    (-r_{31},r_{2}) e^{ - i q_{31} r_{12}} \right ) e^{i q_{2} r_{3}}. \label{boxwavesolsym}
\end{align}

As demonstrated in two-body scattering case in \cite{Guo:2012hv,Guo:2013vsa},  the secular equations or quantization conditions for three-body interaction in finite box is obtained by matching condition, \mbox{$\psi^{(L)}_{sym}  (r_{12},r_{3}; q_{ij}, q_{k}) = \psi_{sym} (r_{12},r_{3}; q_{ij}, q_{k}) $}. All six   plane waves are independent in Eq.(\ref{wavesolsym}) and Eq.(\ref{boxwavesolsym}), therefore,  secular equations are equivalently obtained by matching coefficients of six independent plane waves. To obtain   secular equations, we first consider  the coefficient for a  combination of  \mbox{$\left (e^{iq_{12} r_{12} }-  e^{-iq_{12} r_{12} }  \right )e^{ i q_{3} r_{3}}$}, which is anti-symmetric under exchange of \mbox{$r_{12} \leftrightarrow - r_{12}$} and is obviously forbidden for bosons system. Matching condition for this particular   wave reads
\begin{align}
& A_{sym} (r_{12},r_{3}) -A_{sym} (-r_{12},r_{3})   \nonumber \\
& \quad  \quad \quad  =  A^{(L)}_{sym}  (r_{12},r_{3})- A^{(L)}_{sym}  (-r_{12},r_{3}).
\end{align}
Using Eq.(\ref{coefwavesolsym}) and Eq.(\ref{boxcoefwavesolsym}), matching condition leads to
\begin{align}
& \quad   4   i  \mathbf{ T}_{3}  \left [  \theta(r_{12})  -  \theta(-r_{12})  \right ]   \nonumber \\
& \quad \times  \left [ \frac{  e^{ i (\frac{2}{3} P +q_{3})L}}{1- e^{ i (\frac{2}{3} P +q_{3})L}}   - \frac{e^{2 i  \left (\delta (-q_{23})- \delta (-q_{31})  \right ) }}{   1 - e^{2 i  \left ( \delta (-q_{23}) -\delta (-q_{31})  \right ) }  }  \right ]  \nonumber \\
&  -   4    i  \mathbf{ T}_{1}  \left [  \theta(r_{23}) - \theta(-r_{31})  \right ] \nonumber \\
& \quad \times\left [   \frac{ e^{- i (\frac{2}{3} P + q_{1})L}}{1- e^{- i (\frac{2}{3} P +q_{1})L}}  -    \frac{e^{- 2 i  \left ( \delta (-q_{31}) + \delta (q_{12}) \right ) }}{   1 - e^{- 2 i  \left ( \delta (-q_{31}) + \delta (q_{12}) \right ) }  }  \right ]   \nonumber \\
&  - 4   i  \mathbf{ T}_{2}   \left [  \theta(r_{23}) - \theta(-r_{31})  \right ]   \nonumber \\
& \quad \times \left [    \frac{ e^{ i (\frac{2}{3} P +q_{2})L}}{1- e^{ i (\frac{2}{3} P +q_{2})L}}  - \frac{e^{- 2 i  \left ( \delta (-q_{23}) + \delta (q_{12}) \right ) }}{   1 - e^{- 2 i  \left ( \delta (-q_{23}) + \delta (q_{12}) \right ) }  }  \right ]   =0.
\end{align}
The   matching condition has to be satisfied in all regions in $(r_{12},r_{3})$ plane, see Fig.~\ref{fig:config}, choosing region for an example $(I):$ \mbox{$r_{12}<0$},  \mbox{$r_{23}>0$} and  \mbox{$r_{31}<0$}, thus, we obtain our first secular equation,
\begin{align}
& e^{ i (\frac{2}{3} P +q_{3})L} = e^{2 i  \left (  \delta (-q_{23})-\delta (-q_{31})  \right ) }   . \label{seculareq3}
\end{align}
Other two secular equations are obtained similarly by considering the combination  of \mbox{$\left (e^{i q_{23} r_{12} }-  e^{-iq_{23} r_{12} }  \right )e^{ i q_{1} r_{3}}$} and \mbox{$\left (e^{iq_{31} r_{12} }-  e^{-iq_{31} r_{12} }  \right )e^{ i q_{2} r_{3}}$}  respectively,
\begin{align}
&  e^{ i (\frac{2}{3} P + q_{1})L}= e^{ 2 i  \left ( \delta (-q_{31}) + \delta (q_{12}) \right ) }  , \label{seculareq1}  \\
&  e^{ i (\frac{2}{3} P +q_{2})L} =e^{- 2 i  \left ( \delta (-q_{23}) + \delta (q_{12}) \right ) }  . \label{seculareq2}
\end{align}
Although above three secular equations for three-boson interaction are obtained by choosing a particular region, it is easy to check that the Eqs.(\ref{seculareq3}-\ref{seculareq2}) are indeed the solutions of all six matching conditions in all regions on $(r_{12},r_{3})$ plane. As the matter of fact, the secular equations displayed in Eqs.(\ref{seculareq3}-\ref{seculareq2}) have been obtained long ago by Yang in \cite{Yang:1967bm} as a specific case of N-particle system as \mbox{$N=3$}.  In \cite{Yang:1967bm}, based on Bethe's Hypothesis \cite{Bethe:1931hc,Lieb:1963rt,Yang:1966ty}, {\it i.e.} ``no diffraction'' hypothesis, Yang considered a more general situation of $N$ identical particles problem in one dimension for $\delta$-interaction. Nevertheless,    all three secular equations for three-boson system   appear as  L\"uscher's formula like quantization conditions,
\begin{align}
& \cot (\frac{P}{3} + \frac{q_{3}}{2})L + \cot  \left (  \delta (-q_{31})-\delta (-q_{23})  \right ) =0 , \nonumber \\
&  \cot (\frac{P}{3} + \frac{q_{1}}{2})L + \cot  \left ( - \delta (-q_{31}) - \delta (q_{12}) \right ) =0, \nonumber \\
&  \cot (\frac{P}{3} + \frac{q_{2}}{2})L + \cot  \left ( \delta (-q_{23}) + \delta (q_{12}) \right ) =0 . \label{seculareq}
\end{align}

  \begin{figure}
\begin{center}
\includegraphics[width=3.2 in,angle=0]{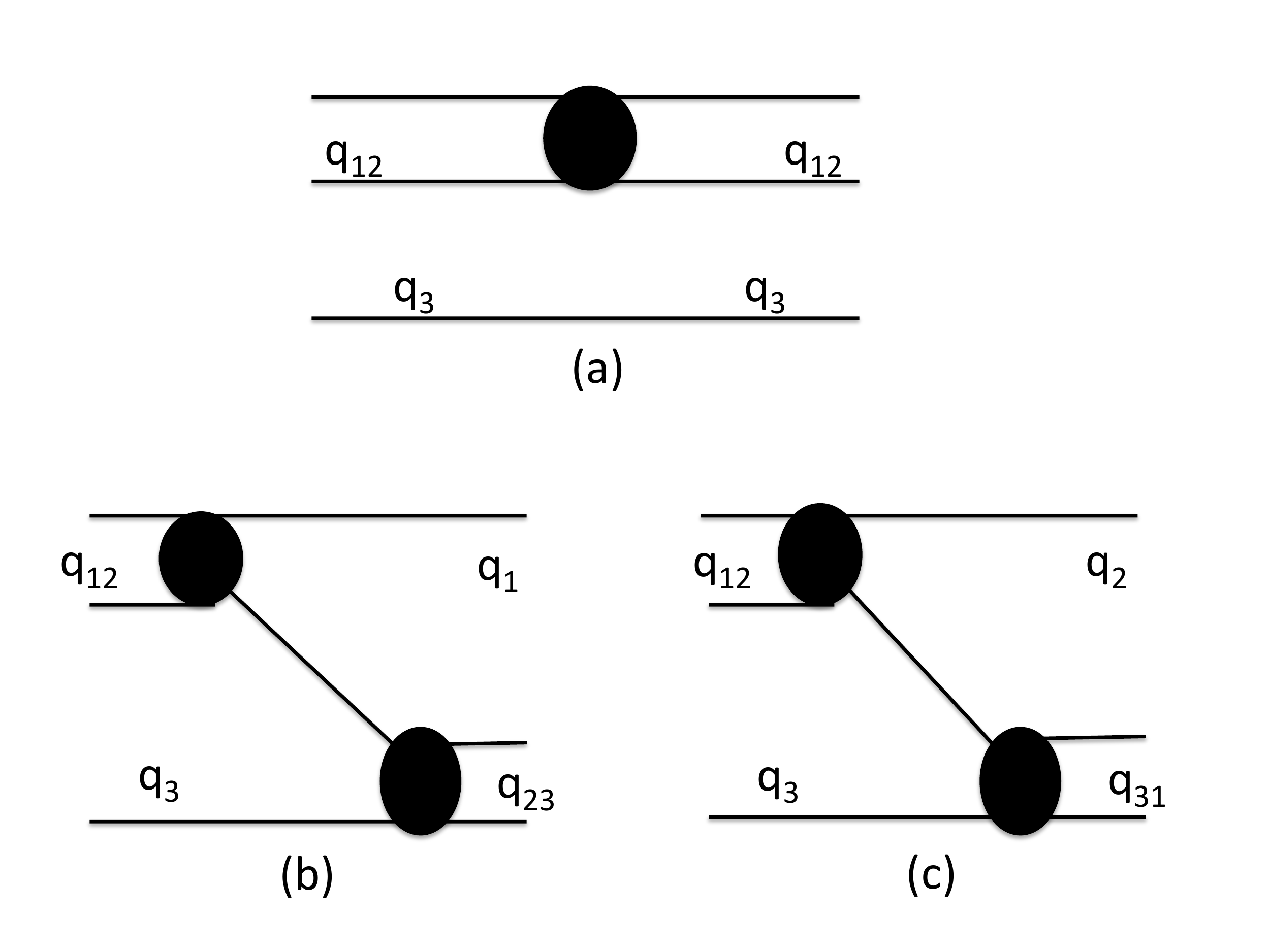}  
\caption{ (a) the disconnected diagram for 3rd particle as a spectator.  (b-c) double scattering contributions from pair $(23)$ and $(31)$ into $(12)$ pair. \label{fig:feynman}  } 
\end{center}
\end{figure}

\section{Discussion and conclusion}\label{summary}
 Using quantization conditions in Eqs.(\ref{seculareq3}-\ref{seculareq2}), we also obtain relations for relative momenta $q_{ij}$, for an example, 
 \begin{align}
 \cot \frac{q_{12} L}{2} + \cot \left ( \delta (q_{12}) + \frac{\delta (-q_{23})+\delta (-q_{31}) }{2} \right) =0. \label{q12quantization}
 \end{align}  
 The $\delta(q_{12})$ comes from the disconnected scattering contribution in $(12)$ pair, see Fig.\ref{fig:feynman}(a), \mbox{$\frac{\delta (-q_{23})+\delta (-q_{31}) }{2} $} is net result of sum over all rescattering contributions from other channels into $(12)$ pair. The physical picture is somehow quite similar to the three-body rescattering effect in three-body decay processes \cite{Khuri:1960zz,Bronzan:1963xn,Aitchison:1965kt,Aitchison:1965zz,Aitchison:1966kt,Pasquier:1968zz,Pasquier:1969dt,Guo:2014vya,Guo:2014mpp,Danilkin:2014cra,Guo:2015kla}. Based on Khuri-Trieman equation approach, the   decay process of a particle $(0)$ into three final particles is described by a sum of all possible decay chains:  $0 \rightarrow (12)3 + 1(23) + (31)2 \rightarrow 123$. For each individual decay chain, the amplitude is product of two-body amplitude and a scalar function that describe the net effect of  three-body rescattering corrections to disconnected two-body contribution. Analogue to rescattering in three-body decay processes, \mbox{$\frac{\delta (-q_{23})+\delta (-q_{31}) }{2} $} may be interpreted as the three-body rescattering corrections to disconnected two-body contribution, $\delta(q_{12})$.

 Assuming that we can treat Faddeev's equation Eq.(\ref{faddeevTeq}) as a perturbation theory, and the leading order solution of Eq.(\ref{faddeevTeq}) is disconnected contribution, 
 \begin{align}
&  T_{(\gamma)}^{(0)} (k; q_{ij} ,q_{k})  \nonumber \\
& \quad  =    (2\pi) \delta( k -q_{\gamma})    (2  \sqrt{\sigma^{2} -\frac{3}{4} q_{\gamma}^{2}} )t_{+}(  \sqrt{\sigma^{2} -\frac{3}{4} q_{\gamma}^{2}} ) .
 \end{align}
 Therefore, the total scattering amplitude \mbox{$T(k_{12}, k_{3}; q_{ij}, q_{k}) = \sum_{\gamma=1}^{3}T_{(\gamma)}^{(0)} (k_{\gamma}; q_{ij} ,q_{k}) $} only has contribution of three disconnected scattering amplitudes, \mbox{$1+2 \rightarrow 1+2$} with particle-3 as a spectator, \mbox{$2+3 \rightarrow 2+3$} with particle-1 as a spectator  and \mbox{$3+1 \rightarrow 3+1$} as particle-2 as a spectator.
 Iterating Eq.(\ref{faddeevTeq}) once, thus, the next-leading order contribution of $T_{(\gamma)}$ is given by,
 \begin{align}
  T_{(\gamma)}^{(1)} & (k; q_{ij} ,q_{k})  = (2  \sqrt{\sigma^{2} -\frac{3}{4} k^{2}} )i t_{+}(  \sqrt{\sigma^{2} -\frac{3}{4} k^{2}} ) \nonumber \\
& \quad \quad \times \sum_{\alpha \neq \gamma} \frac{   (2  \sqrt{\sigma^{2} -\frac{3}{4} q_{\alpha}^{2}} ) i t_{+}(  \sqrt{\sigma^{2} -\frac{3}{4} q_{\alpha}^{2}} ) }{\sigma^{2} - \frac{3}{4} q_{\alpha}^{2} -(k + \frac{q_{\alpha}}{2})^{2} + i \epsilon }.
 \end{align}
   Diagrammatic representation of  $T^{(1)}_{(3)}$ is shown in Fig.\ref{fig:feynman}(b) and Fig.\ref{fig:feynman}(c), which are   double scattering contributions  from pair $(23)$ and $(31)$ into $(12)$ pair.  Now both leading and next-leading oder contributions to  the  on-shell scattering amplitude  $\mathcal{T}_{3}$    are 
   \begin{align}
   \mathcal{T}^{(0)}_{3} + \mathcal{T}^{(1)}_{3} = t_{+}(q_{12}) \left [ 1+ i t_{+}(-q_{23}) +  it_{+} (- q_{31}) \right ]. \label{perturb}
   \end{align}
   In Eq.(\ref{perturb}), the perturbation result of scattering amplitude, $ \mathcal{T}_{3}$,  on right hand side of equation now indeed appears as   the product of disconnected two-body scattering amplitude $t_{+}(q_{12}) $  and the  rescattering corrections to leading  order contribution,  $t_{+}(q_{12}) $.
   
On the other hand,    the asymptotic   behavior of  two-body phase shift  is  given by \mbox{$\delta(q) \rightarrow    -\frac{m V_{0}}{2q} $} as \mbox{$q \rightarrow \infty$},  where $V_{0}>0$ for repulsive interaction and $V_{0}<0$ for attractive interaction. For large $q_{3}$ (the momentum of third particle is well separated from relative momentum of pair $(12)$), thus, \mbox{$q_{23} \rightarrow  \frac{3}{4} q_{3}$} and \mbox{$q_{31} \rightarrow - \frac{3}{4} q_{3}$}, and 
 \begin{align}
& \frac{\delta (-q_{23})+\delta (-q_{31}) }{2} \stackrel{q_{3} \rightarrow \infty}{ \rightarrow } -\frac{m V_{0}}{3} \left ( \frac{1}{q_{3}} - \frac{1}{q_{3}}\right ) =  0,  \nonumber \\
 &\frac{\delta (-q_{23})-\delta (-q_{31}) }{2} \stackrel{q_{3} \rightarrow \infty}{ \rightarrow }     - \frac{2 m V_{0}}{3 q_{3}} \sim \delta (\frac{3 }{4} q_{3}) , \nonumber \\
 &  i t_{+}(-q_{23}) +  it_{+} (- q_{31})\stackrel{q_{3} \rightarrow \infty}{ \rightarrow } \frac{ 2 i m V_{0}}{3} \left ( \frac{1}{q_{3}} - \frac{1}{q_{3}}\right )  =0.
 \end{align}
  Therefore, at large $q_{3}$, the rescattering between a energetic  3rd particle and particles in  pair $(12)$  is less likely to happen. The quantization condition in Eq.(\ref{q12quantization}) and first condition in Eq.(\ref{seculareq})  are thus reduced to isobar model type conditions, \mbox{$ \cot \frac{q_{12} L}{2} + \cot   \delta (q_{12})=0$} and \mbox{$ \cot (\frac{P}{3} + \frac{q_{3}}{2})L + \cot  \delta (\frac{3 }{4} q_{3}) =0$}, in which the rescattering effect from 3rd particle is week and neglected. The reduction of quantization conditions can be understood in follow arguments. Diagrammatically, rescattering amplitudes, such as Fig.\ref{fig:feynman}(b) and Fig.\ref{fig:feynman}(c),  are proportional to propagators \mbox{$\frac{1}{(k- q_{3})(k-q_{1})}$} and \mbox{$\frac{1}{(k- q_{3})(k-q_{2})}$} respectively. When off-shell momentum $q$ is taken close to $q_{3}$,  the amplitude at the pole $k=q_{3}$ position leads to on-shell   scattering amplitude $\mathcal{T}_{3}$ given in Eq.(\ref{phaseshiftT}), meanwhile, the contributions from  Fig.\ref{fig:feynman}(b) and Fig.\ref{fig:feynman}(c)  are proportional to $\frac{1}{2 q_{31}}$ and $\frac{1}{2 q_{23}}$ respectively. Hence, for large $q_{3}$,  rescattering contribution   from channel ($23$) and ($31$) into pair $(12)$   are both highly suppressed by $\frac{1}{q_{3}}$, so that quantization conditions for both $q_{12}$ in Eq.(\ref{q12quantization}) and $q_{3}$ in first condition in Eq.(\ref{seculareq}) are reduced to isobar model like quantization conditions, and  the dominant   contribution is   from disconnected diagram.

Although, McGuire's model display no diffraction effect, our results given in Eq.(\ref{seculareq}) may still hold for a general short range potentia. This may be demonstrated by asymptotic behavior of three-body wave function. The asymptotic form of wave function in one dimension is quite different from that in three dimension, {\it e.g.} two-body scattering wave function in one dimension doesn't fall off in any direction, see  Eq.(\ref{2bwaveasym}). For incoming three free particles, as in three dimension, the one dimensional  three-body wave function also consists of several pieces that  display the different asymptotic behavior and describe different physical processes:  (1)  the contribution from incoming free waves, disconnected diagrams and non-diffracted on-shell rescattering effects all have the form of non-diffraction waves, {\it e.g.} $e^{i q_{ij} r_{12}} e^{ i q_{k} r_{3}}$; (2) the bound state   capture process has the form of, {\it e.g.} $e^{- \chi_{12} |r_{12}|} e^{ i q_{3} r_{3}}$ with a bound state of $(12)$ pair in finial state, which decay exponentially as $(r_{ij},r_{k}) \rightarrow \infty$; (3)  diffraction waves are of the order of, 
\begin{align}
& \int \frac{d k_{12}}{2\pi}  \frac{d k_{3}}{2\pi} \frac{e^{ i k_{12} r_{12}   } e^{ i k_{3}  r_{3} }}{\sigma^{2} - k_{12}^{2} -\frac{3}{4} k_{3}^{2} + i \epsilon}  \nonumber \\
& \quad \quad \quad \stackrel{ (r_{12}, r_{3}) \rightarrow  \infty }{ \propto} \left ( r_{12}^{2}+ \frac{4}{3} r_{3}^{2} \right )^{-\frac{1}{4}} e^{i \sigma  \sqrt{r_{12}^{2}+ \frac{4}{3} r_{3}^{2} }} ,
\end{align}
which describe spherical wave of three-body effect and are suppressed at large distance \cite{Lipszyc:1975rm}. Hence, at large separations of all three particles, the dominant contribution is from  non-diffraction waves. Therefore, we expect that Eq.(\ref{seculareq}) may still hold for a general short range potential.

In summary, McGuire's model is adopted to describe three spinless identical particles scattering in one spatial dimension, we present the  details solutions of Faddeev's equation  for scattering  of three free spinless   particles. The three particles interaction in finite volume is derived in Section \ref{3bbox}.  Our approach of solving three-body interaction in finite volume is a generalization of approach developed in  \cite{Guo:2012hv,Guo:2013vsa} by considering  wave function in configuration representation,  the advantage is that the wave function contains only on-shell   scattering amplitudes.  The quantization conditions by matching wave function in free space and finite volume are   given in terms of two-body scattering phase shifts in Eq.(\ref{seculareq}).   The quantization conditions in   McGuire's model is dramatically simplified due to Bethe's hypothesis, and the quantization conditions   presented in Eq.(\ref{seculareq}) are  L\"uscher's formula like  and are consistent with results obtained in \cite{Yang:1967bm}. Finally, we would like to point out that our results in Eq.(\ref{seculareq}) are presented in terms of two-body scattering phase shift,  although, they are derived based on a particular model, Eq.(\ref{seculareq})  may be more general for pair-wise and short range   interactions. The quantization conditions may be tested   in the near future by one dimensional lattice models, such as  ones studied  in \cite{Gattringer:1992np,Guo:2013vsa}.

\section{ACKNOWLEDGMENTS}
We thank R.~A.~Briceno  for useful discussions.  We also acknowledges support from Department of Physics and Engineering, California State University, Bakersfield, CA.

\appendix

\section{Formal theory of scattering and Faddeev's equation }\label{formalscatt}
\subsection{Formal theory of scattering}
  In the formal theory of scattering    \cite{GellMann:1953zz}, assuming   Hamiltonian of scattering system is given by  the sum of a kinematic term and a interaction term, \mbox{$\hat{H}=\hat{H}_{(0)}+\hat{V}$}, the $S$-matrix is given in terms of the solution of Schr\"odinger equation, 
\begin{equation}
   \langle  f  | \hat{S} | i \rangle =   \langle  f  | \hat{U}(\infty,0 )  \hat{U}(0, -\infty) | i \rangle = \langle\Psi_{f}^{(-)}(0) |\Psi_{i}^{(+)} (0)  \rangle ,  
\end{equation}  
where the unitary operator $\hat{U}$ is given by  \mbox{$\hat{U}(t, t_{0}) =\mathcal{T} \{ \exp \left [- i  \int_{t_{0}}^{t} dt'  e^{i \hat{H}_{(0)} t'  }   \hat{V}  e^{-i \hat{H}_{(0)} t'  }    \right ] \}$}, it has the properties of \mbox{$|\Psi (t) \rangle = \hat{U}(t, t_{0}) | \Psi(t_{0}) \rangle$}. $| \Psi(t) \rangle$ is the solution of time-dependent Schr\"odinger equation, which describes the wave vector of scattering system.  $| i \rangle$ and $| f \rangle$  are initial and final state vectors in the absence of interaction at distant past and future respectively.  The incoming and outgoing wave vectors \mbox{$|  \Psi_{i}^{(+)} (0) \rangle=  \hat{U}(0, -\infty) | i \rangle$} and \mbox{$ \langle \Psi_{f}^{(-)}(0)  | =  \langle  f  | \hat{U}(\infty,0 ) $}      are also given by  Lippmann-Schwinger equation \cite{GellMann:1953zz,Lippmann:1950zz},
\begin{align}
 |  \Psi_{i}^{(+)} (0) \rangle  & =\left [1+ \frac{1}{E_{i} - \hat{H}_{(0)} - \hat{V}+ i \epsilon} \hat{V} \right ]  | i \rangle  \nonumber \\
&=    | i \rangle + \frac{1}{E_{i} - \hat{H}_{(0)} + i \epsilon} \hat{V}    |  \Psi_{i}^{(+)} (0) \rangle,    \label{incomingwave}  \\ 
  \langle \Psi_{f}^{(-)}(0)  |  & =  \langle f | \left [1+ \hat{V} \frac{1}{E_{f} - \hat{H}_{(0)} - \hat{V}+ i \epsilon} \right ] \nonumber \\
 &=  \langle f | +   \langle \Psi_{f}^{(-)}(0)  |   \hat{V} \frac{1}{E_{f} - \hat{H}_{(0)} + i \epsilon}    ,\label{outgoingwave}
\end{align}  
where    $E_{i}$  and $E_{f}$  denote initial and final state energies respectively.

Using Eq.(\ref{outgoingwave}), we first rewrite $S$-matrix to,
\begin{equation}\label{prodamp1}
      \langle f  | \hat{S} | i \rangle   =  \langle  f  |  \Psi_{i}^{(+)} (0) \rangle -  \frac{1}{E_{f}-E_{i}  +i \epsilon }   \langle  f | \hat{T}| i \rangle,
\end{equation}  
where  \mbox{$ \langle  f | \hat{T}| i \rangle =- \langle  f | \hat{V}     |\Psi_{i}^{(+)} (0) \rangle$} is scattering $T$-matrix. With the help of Eq.(\ref{incomingwave}), we obtain the relations for $T$-matrix,
 \begin{equation}
   \hat{T}  =-  \hat{V}   + \hat{V}  \frac{1}{E_{i} - \hat{H} + i \epsilon} \hat{V}   = -  \hat{V}   + \hat{V}  \frac{1}{E_{i} - \hat{H}_{(0)} + i \epsilon} \hat{T}  .
  \end{equation}
In terms of $T$-matrix, the incoming wave reads \mbox{$ |  \Psi_{i}^{(+)} (0) \rangle =\left [1  -\frac{1}{E_{i} - \hat{H}_{(0)} + i \epsilon} \hat{T} \right ] |  i \rangle$}, therefore,
the $S$- and $T$-matrix   are related by
   \begin{equation}\label{smatprod} 
     \langle f  | \hat{S} | i \rangle    =  \langle  f  |  i \rangle + 2 \pi i \delta(E_{i} -E_{f})   \langle  f | \hat{T}| i \rangle .    
\end{equation}

\subsection{Faddeev's equation} \label{faddeevsec}
For three-particle scattering, the   wave vector $|\Psi^{(+)}_{E} \rangle$ satisfies  Schr\"odinger   equation,
\begin{align}
\left ( E- \hat{H}_{(0)} - \hat{V} \right  ) |\Psi^{(+)}_{E}  \rangle =0.
\end{align}
Assuming  pair-wise interactions among each pair of particles, \mbox{$\hat{V}= \sum_{\gamma=1}^{3} \hat{V}_{(\gamma)}  $}, where $\hat{V}_{(\gamma)}$ stands for the pair-wise interaction between $\alpha$-th and $\beta$-th    particles. As shown in \cite{Faddeev:1960su,Faddeev:1965}, the self-consistent equations for three-body wave function depend on the free incoming waves, and  are split into four classes according to four types of asymptotic  free incoming waves: $| i \rangle$ and $| \Phi_{(\gamma)} \rangle$ (\mbox{$\gamma=1,2,3$}), where $| i \rangle$  is solutions of \mbox{$  ( E- \hat{H}_{(0)}      ) | i\rangle =0$} and  represents incoming wave of three free particles,  and  $| \Phi_{(\gamma)} \rangle$ is solution of \mbox{$ ( E- \hat{H}_{(0)} - \hat{V}_{(\gamma)}    ) |\Phi_{(\gamma)} \rangle =0$} and represents free $\gamma$-th particle plus a bound state in $(\alpha \beta)$ pair.

\subsubsection{Scattering of three free particles}
  For  three-body scattering with initial state of free incoming wave $| i \rangle$, the three-body scattering wave vector  has the form of  \mbox{$|\Psi^{(+)}_{E} \rangle = | i \rangle + \sum_{\gamma=1}^{3} | \Psi_{(\gamma)} \rangle $} \cite{Faddeev:1960su,Faddeev:1965},  
where 
 $| \Psi_{(\gamma)} \rangle$      satisfy equation,
\begin{align}
&|\Psi_{(\gamma)} \rangle = \hat{G}_{(\gamma)}  \hat{V}_{(\gamma)} \left ( |  i \rangle + | \Psi_{(\alpha)} \rangle+ | \Psi_{(\beta)} \rangle\right ), \ \ \alpha \neq \beta \neq \gamma. \label{psik}
\end{align}
The Green's function \mbox{$\hat{G}_{(\gamma) }=   (E-\hat{H}_{(0)}-\hat{ V}_{(\gamma)} + i \epsilon   )^{-1}$}  is solution of equation, \mbox{$\left ( E- \hat{H}_{(0)} - \hat{V}_{(\gamma)} \right  ) \hat{G}_{(\gamma) } = 1$}. Green's function $\hat{G}_{(\gamma) } $ is related to two-body scattering amplitude by, \mbox{$\hat{G}_{(\gamma) } =\hat{G}_{(0)} \left ( 1  -  \hat{t}_{(\gamma)} \hat{G}_{(0)} \right)  $}, where \mbox{$  \hat{G}_{(0)}  = (E-\hat{H}_{(0)} + i \epsilon)^{-1} $}  and  \mbox{$\hat{t}_{(\gamma)} =- \hat{V}_{(\gamma)} + \hat{V}_{(\gamma)} \hat{G}_{(0)}   \hat{t}_{(\gamma)}   $} are free Green's function and  two-body scattering $T$-matrix  in $(\alpha \beta)$ pair channel respectively. Therefore, we   found relations \mbox{$\hat{V}_{(\gamma)} \hat{G}_{(\gamma) } = -  \hat{t}_{(\gamma)} \hat{G}_{(0)} $},  and
\begin{align}
&  \hat{V}_{(\gamma)} |\Psi_{(\gamma)} \rangle =-   \hat{t}_{(\gamma)}   \hat{G}_{(0)}  \hat{V}_{(\gamma)} \left ( |  i \rangle + | \Psi_{(\alpha)} \rangle+ | \Psi_{(\beta)} \rangle\right ). \label{psikagain}
\end{align}

The total three-body scattering amplitude is given by \mbox{$\hat{T} | i \rangle=- \hat{V} |\Psi^{(+)}_{E}  \rangle= \sum_{\gamma=1}^{3} \hat{T}_{(\gamma)}| i \rangle$} \cite{Faddeev:1960su,Faddeev:1965},  where
\begin{align}
\hat{T}_{(\gamma)}| i \rangle =-\hat{V}_{(\gamma)} |\Psi^{(+)}_{E}  \rangle.
\end{align}
 Using Eq.(\ref{psikagain}), we thus have 
 \begin{align}
&\hat{T}_{(\gamma)}  | i \rangle 
=-\left (1-\hat{t}_{(\gamma)} \hat{G}_{(0)} \right)\hat{V}_{(\gamma)}  \left ( |  i \rangle + | \Psi_{(\alpha)} \rangle+ | \Psi_{(\beta)} \rangle\right )  , \label{Tgop}   \\
&| \Psi_{(\gamma)} \rangle = - \hat{G}_{(0)} \hat{T}_{(\gamma)}  |  i \rangle . \label{psiTgamma}
\end{align}
Eq.(\ref{Tgop}) and Eq.(\ref{psiTgamma}) together lead to the well-known   Faddeev's equation  for three particles scattering \cite{Faddeev:1960su,Faddeev:1965},
\begin{align}
\hat{T}_{(\gamma)} &  =   \hat{t}_{(\gamma)}   - \hat{t}_{(\gamma)} \hat{G}_{(0)}   \left ( \hat{T}_{(\alpha)} + \hat{T}_{(\beta)} \right), \ \ \alpha \neq \beta \neq \gamma. \label{fdopeq}
\end{align}

\subsubsection{Scattering by a bound state}
For  the case of $i$-the particle incident on a bound state of other two particles pair, the  initial state of free incoming wave is given by \mbox{$| \Phi_{(i)} \rangle $}. The three-body wave vector  has the form of  \mbox{$|\Psi^{(+)}_{E} \rangle =  \sum_{\gamma=1}^{3} | \Psi_{(\gamma)} \rangle $} \cite{Faddeev:1960su,Faddeev:1965},  
where 
 $| \Psi_{(\gamma)} \rangle$      satisfy equation,
\begin{align}
|\Psi_{(\gamma)} \rangle = \delta_{\gamma, i} | \Phi_{(i)} \rangle+\hat{G}_{(\gamma)}  \hat{V}_{(\gamma)}  & \left (  | \Psi_{(\alpha)} \rangle + | \Psi_{(\beta)} \rangle\right ),  \nonumber \\
&  \alpha \neq \beta \neq \gamma. \label{boundpsik}
\end{align}
The    total scattering amplitude for a particle scattering with a bound state is   given by \mbox{$\hat{T} | \Phi_{(i)} \rangle  =- \hat{V} |\Psi^{(+)}_{E}  \rangle= \sum_{\gamma=1}^{3} \hat{T}_{(\gamma)} | \Phi_{(i)} \rangle  $} \cite{Faddeev:1960su,Faddeev:1965},  where 
 \begin{align}
& \hat{T}_{(\gamma)} | \Phi_{(i)} \rangle  =-\hat{V}_{(\gamma)} |\Psi^{(+)}_{E}  \rangle.
\end{align}
Thus, we find
\begin{align}
& \hat{T}_{(\gamma)} | \Phi_{(i)} \rangle   =- \delta_{\gamma, i} \hat{V}_{(\gamma)} | \Phi_{(i)} \rangle
    + \hat{t}_{(\gamma)}  \left (   | \Psi_{(\alpha)} \rangle+ | \Psi_{(\beta)} \rangle\right )   , \\
& | \Psi_{(\gamma)} \rangle = - \hat{G}_{(0)} \hat{T}_{(\gamma)}  | \Phi_{(i)} \rangle  .
\end{align}
 The  Faddeev's equation  for $i$-the particle incident on a bound state of other two particles pair yields
\begin{align}
\hat{T}_{(\gamma)}   =- \delta_{\gamma, i} \hat{V}_{(\gamma)}   - \hat{t}_{(\gamma)} \hat{G}_{(0)}  & \left ( \hat{T}_{(\alpha)} + \hat{T}_{(\beta)} \right),   \ \  \alpha \neq \beta \neq \gamma. \label{boundfdopeq}
\end{align}

\section{Solutions of Faddeev's equation for short range interaction in free space}\label{TsolFaddeev}

In this section, we  first show the details of solution of Faddeev's equation, Eq.(\ref{faddeevTeq}),
\begin{align}
&T_{(\gamma)} (k ; q_{ij}, q_{k})   =  (2\sqrt{\sigma^{2}    -\frac{3}{4} k^{2} }) i t_{+}(\sqrt{\sigma^{2}    -\frac{3}{4} k^{2} } )    \nonumber \\
&  \quad   \times \left [ \frac{  v_{(\gamma)} (k; q_{ij}, q_{k}) }{ i m V_{0}}   \right. \nonumber \\
& \quad\quad \quad\quad \left.    + \quad   i \int_{-\infty}^{\infty}   \frac{d q }{2\pi}     \frac{  T_{(\alpha)}  (q; q_{ij}, q_{k})  + T_{(\beta)} (q; q_{ij}, q_{k})    }{ \sigma^{2}     -\frac{3}{4} q^{2} - (k  + \frac{q}{2})^{2}  + i \epsilon}     \right ], \nonumber \\
&\quad\quad \quad\quad \quad\quad \quad\quad  \quad\quad \quad\quad \quad  \quad\quad   \alpha \neq \beta \neq \gamma , \nonumber
\end{align}
where
\begin{align}
& v_{(\gamma)} ( k ;q_{ij}, q_{k})  = \int_{-\infty}^{\infty}  d r_{\alpha \beta} d r_{\gamma}  e^{-i k  r_{\gamma}}   \nonumber \\
& \quad \quad \quad \quad   \quad \quad  \times m V_{0} \delta (r_{\alpha \beta})   \psi_{(0)} (r_{\alpha \beta}, r_{\gamma};q_{ij}, q_{k})  .   \nonumber
\end{align}
Then, using the solutions obtained by solving Faddeev's equation, we demonstrate how the three-body scattering wave function is constructed.

\subsection{Solution of $T$ amplitudes}\label{Tsolindv}

  Let's first consider  a free incoming wave,
\begin{equation}
\psi_{(0)} =e^{i q_{12} r_{12}} e^{ i q_{3} r_{3}}=e^{i q_{23} r_{23}} e^{ i q_{1} r_{1}}=e^{i q_{31} r_{31}} e^{ i q_{2} r_{2}}, \label{psi0}
\end{equation}
therefore, 
\begin{align}
v_{(\gamma)} (k; q_{12}, q_{3})= m V_{0} (2\pi) \delta(k-q_{\gamma}),  \ \  \gamma =1,2,3.
\end{align}

First of all, let's introduce three new  functions,   
\begin{align}
Z(k) &= \sum_{\gamma=1}^{3} T_{(\gamma) } (k; q_{ij} , q_{k}) , \nonumber \\
X(k) &=T_{(3)}(k; q_{ij} , q_{k} ) - T_{(1)}(k; q_{ij} , q_{k} )  , \nonumber \\
Y(k) &=T_{(3)}(k; q_{ij} , q_{k} ) - T_{(2)}(k; q_{ij} , q_{k} ) , 
\end{align}
 thus,  Faddeev's equation, Eq.(\ref{faddeevTeq}), can be re-expressed as three decoupled integral equations for $(X,Y,Z)$ functions,
 \begin{align}
& \frac{1}{  (2\sqrt{ \sigma^{2}   -\frac{3}{4} k^{2} }) i t_{+}(\sqrt{\sigma^{2}   -\frac{3}{4} k^{2} } ) }Z (k )     \nonumber \\
&  \quad \quad  \quad \quad   =  - i (2\pi) \sum_{\gamma=1}^{3} \delta(k - q_{\gamma})  \nonumber \\
&\quad \quad  \quad \quad   +    2 i \int_{-\infty}^{\infty}   \frac{d q }{2\pi}     \frac{ Z  (q )    }{ \sigma^{2}  -\frac{3}{4} q^{2} - (k  + \frac{q}{2})^{2}  + i \epsilon}     ,   \label{faddeevZ}
\end{align}
 \begin{align}
& \frac{1}{  (2\sqrt{ \sigma^{2}   -\frac{3}{4} k^{2} }) i t_{+}(\sqrt{\sigma^{2}   -\frac{3}{4} k^{2} } ) }X (k )     \nonumber \\
& \quad \quad  \quad \quad  =  - i (2\pi)  \delta(k - q_{3}) + i (2\pi)  \delta(k - q_{1}) \nonumber \\
&  \quad \quad   \quad \quad \quad - i \int_{-\infty}^{\infty}   \frac{d q }{2\pi}     \frac{ X  (q )    }{ \sigma^{2}-\frac{3}{4} q^{2} - (k  + \frac{q}{2})^{2} + i \epsilon }     ,  \label{faddeevX}
\end{align}
and
 \begin{align}
& \frac{1}{  (2\sqrt{ \sigma^{2}  -\frac{3}{4} k^{2} }) i t_{+}(\sqrt{\sigma^{2}  -\frac{3}{4} k^{2} } ) }Y (k )     \nonumber \\
& \quad \quad  \quad \quad  =  - i (2\pi)  \delta(k - q_{3}) + i (2\pi)  \delta(k - q_{2}) \nonumber \\
&  \quad \quad   \quad \quad \quad - i \int_{-\infty}^{\infty}   \frac{d q }{2\pi}     \frac{ Y  (q )    }{ \sigma^{2}  -\frac{3}{4} q^{2} - (k  + \frac{q}{2})^{2}  + i \epsilon}     . \label{faddeevY}
\end{align}

Next, let's  solve Eq.(\ref{faddeevZ}) first. According to  \cite{McGuire:1964zt}, three-body problem with  equal-strength $\delta$-function potentials is exactly solvable, diffraction effects are cancelled out, the solution of  wave function is expressed as sum of six possible plane waves, see Eq.(\ref{mcguirewave}). Therefore, the three-body scattering amplitudes can only be given by the sum of pole terms, see Eq.(\ref{Tsolmcguire}). The   strategy of solving Eq.(\ref{faddeevZ}-\ref{faddeevY}) is thus  to make an ansatz of solution as the sum of six possible pole terms,  the pole positions are given in terms of the momenta of incoming wave. Each pole term is then   assigned with a constant coefficient. While the ansatz of solution is plugged into integral equations Eq.(\ref{faddeevZ}-\ref{faddeevY}), by carefully defining the integration of contour and also requiring that the branch cut contributions on both sides have to be cancelled out as the consequence of Bethe's hypothesis, then, the coefficients of pole terms can be fixed by matching both sides of equations.

In follows, we show how the Eq.(\ref{faddeevZ}) is satisfied by the ansatz,
\begin{align}
 Z(k) &=( 2 \pi i ) \sum_{\gamma=1}^{3} \kappa_{\gamma}  \delta(k - q_{\gamma})  \nonumber \\
&+  \frac{\lambda k }{(k - q_{3} - i \epsilon)  (k - q_{2} - i \epsilon)(k - q_{1} + i \epsilon)     }.
\end{align}
Instead of deforming contour of integration in Eq.(\ref{faddeevZ}), equivalently, we will adopt $i \epsilon$ prescription in this work, and assign the small imaginary parts to relative momenta to avoid poles    on real axis. The left hand side of Eq.(\ref{faddeevZ}) is thus given by
 \begin{align}
&LHS =  \sum_{\gamma =1}^{3}  \frac{\kappa_{\gamma}   }{  (2\sqrt{ q_{\alpha \beta}^{2}}) i t_{+}(\sqrt{q_{\alpha \beta}^{2}} ) }( 2 \pi i )\delta(k - q_{\gamma})  \nonumber \\
& \quad \quad -    \frac{  \lambda \left (\frac{1}{ i m V_{0}} + \frac{1}{2 \sqrt{ \sigma^{2}  -\frac{3}{4} k^{2} }} \right ) k }{(k - q_{3} - i \epsilon)  (k - q_{2} - i \epsilon)(k - q_{1} + i \epsilon)     } ,    \label{LHSZ}
\end{align}
where \mbox{$\alpha \neq \beta \neq \gamma$}.
The integration on right hand side of  Eq.(\ref{faddeevZ}) is carried out by  closing the contour in upper half plane and picking up poles, \mbox{$q = -\frac{k}{2}+ \sqrt{\sigma^{2} - \frac{3}{4} k^{2}} + i \epsilon$}, \mbox{$q_{3} + i \epsilon$} and \mbox{$q_{2} + i \epsilon$},  thus, we find
\begin{align}
 & RHS  =  - i  \sum_{\gamma=1}^{3} (2\pi) \delta(k - q_{\gamma})  \nonumber \\
 &+    2   \sum_{\gamma=1}^{3}   \frac{ \kappa_{\gamma}    }{ \left (k  + \frac{q_{\gamma}}{2}  + \sqrt{q_{\alpha \beta}^{2} } + i \epsilon \right ) \left (k  + \frac{q_{\gamma}}{2}  - \sqrt{q_{\alpha \beta}^{2} } - i \epsilon \right )  }   \nonumber \\
 &+       \frac{\lambda \left ( 1 - \frac{ k}{2 \sqrt{\sigma^{2} - \frac{3}{4}k^{2}} } \right )  }{(k - q_{3} - i \epsilon)  (k - q_{2} - i \epsilon)(k - q_{1} + i \epsilon)        } \nonumber \\
 & - \frac{ \lambda \frac{q_{3}}{2 q_{23}  q_{31}}}{ \left ( k + \frac{q_{3}}{2} + \sqrt{q_{12}^{2}} + i \epsilon \right )  \left ( k + \frac{q_{3}}{2} - \sqrt{q_{12}^{2}} - i \epsilon \right ) }  \nonumber \\
  & - \frac{ \lambda \frac{q_{2}}{2 q_{12}  q_{23}}}{ \left ( k + \frac{q_{2}}{2} + \sqrt{q_{31}^{2}} + i \epsilon \right )  \left ( k + \frac{q_{2}}{2} - \sqrt{q_{31}^{2}} - i \epsilon \right ) } .
\end{align}
We can clearly see that the branch cut contribution, the terms proportional to  \mbox{$\frac{ 1}{ \sqrt{\sigma^{2} - \frac{3}{4}k^{2}} }$}, on both sides of Eq.(\ref{faddeevZ}) cancel out completely.  Next, the square root terms,  $\sqrt{q_{\alpha \beta}^{2}}$,  are  handled by assigning a small imaginary part to \mbox{$q_{12} \rightarrow q_{12} + i 0^{+}$},   the imaginary part for \mbox{$q_{23} \rightarrow q_{23} - i 0^{+}$} and \mbox{$q_{31}\rightarrow q_{31} - i 0^{+}$} are determined completely by relations, \mbox{$q_{23} = -\frac{1}{2}q_{12} + \frac{3}{4} q_{3}$} and \mbox{$q_{31} = -\frac{1}{2}q_{12} - \frac{3}{4} q_{3}$} respectively.   In addition, our convention for complex square root is given by $\sqrt{q^{2} \pm i 0^{+}} = \pm \sqrt{q^{2}}$, therefore, \mbox{$\sqrt{(q \pm i 0^{+})^{2}} = \sqrt{q^{2} \pm 2 q i 0^{+}}  = \pm q$}. Thus,  with our assignment of imaginary part to $q_{12}$, we obtain relations, \mbox{$\sqrt{ (q_{12}+ i 0^{+})^{2}} = q_{12}$}, \mbox{$\sqrt{ (q_{23}- i 0^{+})^{2}} = -q_{23}$} and \mbox{$\sqrt{ (q_{31} - i 0^{+})^{2}} = - q_{31}$}. Hence,  the right hand side of Eq.(\ref{faddeevZ})  now can be reexpressed by
\begin{align}
 & RHS  =  -  \sum_{\gamma=1}^{3} (2\pi i) \delta(k - q_{\gamma})  -  (2\pi i) \delta(k-q_{3}) \frac{ \kappa_{1}}{q_{23}} \nonumber \\
 &    + 2    \frac{ \sum_{\gamma=1}^{3}   \kappa_{\gamma}  (k - q_{\gamma})  }{(k - q_{3} - i \epsilon)  (k - q_{2} - i \epsilon)(k - q_{1} + i \epsilon)   }    \nonumber \\
 &+       \frac{\lambda \left ( 1 - \frac{ k}{2 \sqrt{\sigma^{2} - \frac{3}{4}k^{2}} }    -  \frac{q_{3}  (k - q_{3})  }{2 q_{23}  q_{31}} - \frac{q_{2}(k -q_{2})  }{2 q_{12}  q_{23}}  \right )  }{(k - q_{3} - i \epsilon)  (k - q_{2} - i \epsilon)(k - q_{1} + i \epsilon)        }   .  \label{RHSZ}
\end{align}
Comparing Eq.(\ref{LHSZ}) to Eq.(\ref{RHSZ}),   the branch cut is cancelled out, and the coefficients are given by
\begin{align}
& \kappa_{1} = \frac{i m V_{0}}{1 - \frac{i m V_{0}}{2 q_{23}}}, \ \ \kappa_{2} = \frac{i m V_{0}}{1 - \frac{i m V_{0}}{2 q_{31}}},  \nonumber \\
& \kappa_{3} = \frac{i m V_{0}  \left ( 1+ \frac{i m V_{0}}{2 q_{23}} \right )}{ \left ( 1 + \frac{i m V_{0}}{2 q_{12}} \right ) \left ( 1 - \frac{i m V_{0}}{2 q_{23}} \right )},  \nonumber \\
& \lambda = - \frac{6 ( i m V_{0} )^{2} }{ \left ( 1 + \frac{i m V_{0}}{2 q_{12}} \right ) \left ( 1 - \frac{i m V_{0}}{2 q_{23}} \right )\left ( 1 - \frac{i m V_{0}}{2 q_{31}} \right )}.
\end{align}

  The solutions of Eq.(\ref{faddeevX}-\ref{faddeevY}) are found in a similar way,
\begin{align}
 X  ( k ) &=   (2\pi i ) \delta(k-q_{3}) \frac{   i m V_{0}  }{ \left (1+ \frac{ i  m V_{0}}{2 q_{12}} \right )\left (1- \frac{ i  m V_{0}}{2 q_{23}} \right )  }      \nonumber \\
 &  -  (2 \pi i )\delta(k-q_{1})   \frac{  i m V_{0}    }{  1- \frac{ i  m V_{0}}{2 q_{23}}     }    \nonumber \\
 &    +  \frac{ \frac{\left ( 2 q_{31} \right ) \left ( i m V_{0}  \right )^{2} }{ \left (1+ \frac{ i  m V_{0}}{2 q_{12}} \right )\left (1- \frac{ i  m V_{0}}{2 q_{23}} \right )  } }{\left ( k- q_{3} -i \epsilon  \right ) \left (k- q_{2} - i \epsilon  \right )  \left (k- q_{1} + i \epsilon  \right ) }  ,
\end{align}
and
 \begin{align}
Y  ( k ) &=         (2 \pi i ) \delta(k-q_{3})   \frac{  i m V_{0}     }{  1+ \frac{ i  m V_{0}}{2 q_{12}}    }   \nonumber \\
& -( 2 \pi i ) \delta(k-q_{2})   \frac{   i m V_{0}    }{1- \frac{ i  m V_{0}}{2 q_{31}}}    \nonumber \\
&- \frac{   \frac{ \left (  2 q_{23} \right ) \left (   i  m V_{0}  \right )^{2} }{ \left (1+ \frac{ i  m V_{0}}{2 q_{12}} \right )\left (1- \frac{ i  m V_{0}}{2 q_{31}} \right )  }  }{ \left ( k - q_{3} -i \epsilon  \right ) \left (k- q_{2} - i \epsilon  \right )  \left (k- q_{1} + i \epsilon  \right ) }        .  
\end{align}

In the end, the solutions of $T_{(1,2,3)}$ for free incoming wave \mbox{$\psi_{(0)}=e^{i q_{12} r_{12}} e^{ i q_{3} r_{3}}$} are
 \begin{align}
 T_{(3)}  &( k; q_{ij} ,q_{k} ) =   (2 \pi i ) \delta(k-q_{3})  \frac{    i m V_{0}     }{\left (1+ \frac{ i  m V_{0}}{2 q_{12}} \right ) \left (1- \frac{ i  m V_{0}}{2 q_{23}} \right ) }  \nonumber \\
 &      -   \frac{   \frac{    \left (  i m V_{0}   \right )^{2}  \left ( 2k + q_{3} \right )   }{\left (1+ \frac{ i  m V_{0}}{2 q_{12}} \right ) \left (1- \frac{ i  m V_{0}}{2 q_{23}} \right ) \left (1- \frac{ i  m V_{0}}{2 q_{31}} \right ) } }{ \left ( k- q_{3} -i \epsilon  \right ) \left (k - q_{2} - i \epsilon  \right )  \left (k- q_{1} + i \epsilon  \right )}                  ,   \label{T3sol}
 \end{align} 
  \begin{align}
 T_{(1)}&  ( k ; q_{ij}, q_{k} ) =  (2 \pi i ) \delta(k -q_{1})   \frac{   i m V_{0}   }{ 1- \frac{ i  m V_{0}}{2 q_{23}}     }  \nonumber \\
 &   -   \frac{   \frac{    \left (  i m V_{0}   \right )^{2}  \left (2k + q_{1} - i m V_{0} \right ) }{\left (1+ \frac{ i  m V_{0}}{2 q_{12}} \right ) \left (1- \frac{ i  m V_{0}}{2 q_{23}} \right ) \left (1- \frac{ i  m V_{0}}{2 q_{31}} \right ) }  }{ \left ( k- q_{3} -i \epsilon  \right ) \left (k- q_{2} - i \epsilon  \right )  \left (k- q_{1} + i \epsilon  \right )}           , \label{T1sol}
  \end{align}
  \begin{align}
 T_{(2)}  &( k; q_{ij} ,q_{k}  ) = (2 \pi i ) \delta(k-q_{2})   \frac{  i m V_{0}   }{  1- \frac{ i  m V_{0}}{2 q_{31}}  } \nonumber \\
 &           + (2 \pi i ) \delta(k-q_{3})  \frac{    i m V_{0}    \left ( \frac{ i  m V_{0}}{2 q_{23}} \right )   }{\left (1+ \frac{ i  m V_{0}}{2 q_{12}} \right ) \left (1- \frac{ i  m V_{0}}{2 q_{23}} \right ) }   \nonumber \\
 &    -   \frac{  \frac{    \left (  i m V_{0}   \right )^{2} \left ( 2k + q_{2} + i m V_{0}  \right )   }{\left (1+ \frac{ i  m V_{0}}{2 q_{12}} \right ) \left (1- \frac{ i  m V_{0}}{2 q_{23}} \right ) \left (1- \frac{ i  m V_{0}}{2 q_{31}} \right ) }}{ \left ( k- q_{3} -i \epsilon  \right ) \left (k- q_{2} - i \epsilon  \right )  \left (k- q_{1} + i \epsilon  \right )}        .   \label{T2sol}
\end{align} 
The total   three-body  scattering amplitude, \mbox{$T(k_{\alpha \beta}, k_{\gamma} ; q_{ij}, q_{k})$}, is determined by  Eq.(\ref{3bTsum}).   As the consequence of Bethe's hypothesis,   the physical scattering process for equal-strength $\delta$-function potential and equal mass particles do not create any new momenta, see \cite{McGuire:1964zt}. The final relative momenta in any pair configuration, for instance \mbox{$(k_{12},k_{3})$}, can only be \mbox{$(\pm q_{ij}, q_{k})$} where \mbox{$k=1,2,3$} and \mbox{$i\neq j \neq k$}. Therefore, we may define  on-shell $S$-matrix by
 \begin{align}
 & (2\pi) \delta( k_{12} - q_{ij} ) (2\pi) \delta(k_{3} - q_{k}) S(k_{12},k_{3} ) \nonumber \\
 & \quad  =  (2\pi) \delta( k_{12} - q_{12} ) (2\pi) \delta(k_{3} - q_{3}) \nonumber \\
 & \quad + (2\pi i) \delta(\sigma^{2} - k_{12}^{2} - \frac{3}{4} k^{3}_{3} ) T(k_{12},k_{3}; q_{ij}, q_{k}) .
\end{align} 
For free incoming    wave \mbox{$\psi_{(0)}=e^{i q_{12} r_{12}} e^{ i q_{3} r_{3}}$},  six possible on-shell $S$-matrix elements  are
 \begin{align}
&  \left ( S(q_{12},q_{3} )  , S(-q_{12},q_{3} ) , S(q_{23},q_{1} ) , S(-q_{23},q_{1} ) , \right. \nonumber \\
 & \quad\quad\quad\quad\quad\quad\quad\quad\quad\quad   \left.  S(q_{31},q_{2} )  ,   S(-q_{31},q_{2} ) \right ) \nonumber \\
 & \quad \quad\quad \quad\quad \quad \quad \quad    = (s_{1},s_{2},s_{3},s_{4},s_{5},s_{6}), 
 \end{align}
 where
  \begin{align}
s_{1}&= \frac{    1  }{\left (1+ \frac{ i  m V_{0}}{2 q_{12}} \right ) \left (1- \frac{ i  m V_{0}}{2 q_{23}} \right ) \left (1- \frac{ i  m V_{0}}{2 q_{31}} \right ) }    ,  \nonumber \\
 s_{2}&= \frac{ \left (- \frac{ i  m V_{0}}{2 q_{12}} \right )  \left [ 1 +  \left ( \frac{ i  m V_{0}}{2 q_{23}} \right ) \left ( \frac{ i  m V_{0}}{2 q_{31}} \right )  \right ]  }{\left (1+ \frac{ i  m V_{0}}{2 q_{12}} \right ) \left (1- \frac{ i  m V_{0}}{2 q_{23}} \right ) \left (1- \frac{ i  m V_{0}}{2 q_{31}} \right ) }  ,   \nonumber \\ 
    s_{3}&= \frac{    \left ( \frac{ i  m V_{0}}{2 q_{23}} \right ) \left ( \frac{ i  m V_{0}}{2 q_{31}} \right )   }{\left (1+ \frac{ i  m V_{0}}{2 q_{12}} \right ) \left (1- \frac{ i  m V_{0}}{2 q_{23}} \right ) \left (1- \frac{ i  m V_{0}}{2 q_{31}} \right ) }  ,  \nonumber \\
 s_{4} &= \frac{     \left (  \frac{ i  m V_{0}}{2 q_{31}} \right )   }{\left (1+ \frac{ i  m V_{0}}{2 q_{12}} \right ) \left (1- \frac{ i  m V_{0}}{2 q_{23}} \right ) \left (1- \frac{ i  m V_{0}}{2 q_{31}} \right ) } ,   \nonumber \\  s_{5} & = s_{3}, \nonumber \\
 s_{6} &= \frac{     \left (  \frac{ i  m V_{0}}{2 q_{23}} \right )   }{\left (1+ \frac{ i  m V_{0}}{2 q_{12}} \right ) \left (1- \frac{ i  m V_{0}}{2 q_{23}} \right ) \left (1- \frac{ i  m V_{0}}{2 q_{31}} \right ) }   .
 \end{align}

The  solutions of $T$ amplitudes of Faddeev equation and $S$-matrix for rest of five independent free incoming   waves can be obtained from solutions given in Eq.(\ref{T3sol}-\ref{T2sol}) by relabelling sub-indices. 

(1) for \mbox{$\psi_{(0)}=e^{- i q_{12} r_{12}} e^{ i q_{3} r_{3}}$}, solutions of $T$ amplitudes  are given by \mbox{$T_{(1)} \leftrightarrow T_{(2)}$}, $T_{(3)}$ remains same. The $S$-matrix elements are   \mbox{$(s_{2},s_{1},s_{4},s_{3},s_{6},s_{5})$};

 (2) for \mbox{$\psi_{(0)}=e^{ i q_{23} r_{12}} e^{ i q_{1} r_{3}}$}, solutions of $T$ amplitudes  are given by \mbox{$T_{(3)} \rightarrow T_{(2)}$}, \mbox{$T_{(1)} \rightarrow T_{(3)}$},  and   \mbox{$T_{(2)} \rightarrow T_{(1)}$}. The $S$-matrix elements are   \mbox{$(s_{5},s_{4},s_{1},s_{6},s_{3},s_{2})$}.  
 
 (3) for \mbox{$\psi_{(0)}=e^{- i q_{23} r_{12}} e^{ i q_{1} r_{3}}$}, solutions of $T$ amplitudes  are given by \mbox{$T_{(1)} \leftrightarrow T_{(3)}$}, $T_{(2)}$ remains same. The $S$-matrix elements are  \mbox{$(s_{4},s_{5},s_{6},s_{1},s_{2},s_{3})$}; 
 
 (4) for \mbox{$\psi_{(0)}=e^{ i q_{31} r_{12}} e^{ i q_{2} r_{3}}$}, solutions of  $T$ amplitudes  are given by \mbox{$T_{(3)} \rightarrow T_{(1)}$}, \mbox{$T_{(1)} \rightarrow T_{(2)}$},  and   \mbox{$T_{(2)} \rightarrow T_{(3)}$}. The $S$-matrix elements are   \mbox{$(s_{3},s_{6},s_{5},s_{2},s_{1},s_{4})$};   
 
 (5) for \mbox{$\psi_{(0)}=e^{- i q_{31} r_{12}} e^{ i q_{2} r_{3}}$}, solutions of $T$ amplitudes  are given by \mbox{$T_{(3)} \leftrightarrow T_{(2)}$}, $T_{(1)}$ remains same. The $S$-matrix elements are   \mbox{$(s_{6},s_{3},s_{2},s_{5},s_{4},s_{1})$}.

In the end of this subsection, we also like to point out that the choice of imaginary part assignment for complex square root is not unique, for instance, we could   assign a small imaginary part to $q_{23}$ instead of $q_{12}$. If so, the solutions obtained by assigning $i \epsilon$ to $q_{23}$  are equivalent to relabel   particles numbers  by  \mbox{$1 \rightarrow 2$}, \mbox{$2 \rightarrow 3$} and \mbox{$3 \rightarrow 1$} from the solutions obtained by assigning $i \epsilon$ to $q_{12}$.

\subsection{Construction of wave function from solution of $T$'s}\label{constwave}
With the solutions of scattering amplitudes in Eq.(\ref{T3sol}-\ref{T2sol}), we are now at the position of constructing wave function of three-body scattering. We show some details of wave function construction in this section for the   incoming wave \mbox{$\psi_{(0)}=e^{i q_{12} r_{12}} e^{ i q_{3} r_{3}}$}  as an example. Using   Eq.(\ref{waveisobar}), we thus obtain 
\begin{align}
 & \psi    (r_{12}, r_{3};q_{ij}, q_{k})  =  e^{i q_{12} r_{12}} e^{ i q_{3} r_{3}}  \nonumber   \\
&  +   \int_{-\infty}^{\infty}   \frac{d k_{12} }{2\pi}   \frac{d k_{3}}{2\pi}      \frac{ e^{i   k_{12} r_{12}  }e^{i k_{3}  r_{3}  }  }{   k_{12}^{2}  + \frac{3}{4} k_{3}^{2} -  \sigma^{2} - i \epsilon } \nonumber \\
& \quad  \times \left [    T_{(3)}  ( k_{3};q_{ij}, q_{k}) + T_{(1)}  (k_{1};q_{ij}, q_{k})+ T_{(2)}  (k_{1};q_{ij}, q_{k})    \right ]    , \label{reconwave}
\end{align}
where \mbox{$k_{1} =- k_{12} - \frac{ k_{3} }{2} $} and  \mbox{$k_{2} = k_{12} - \frac{ k_{3} }{2} $}. For each individual $\psi_{(1,2,3)}$, see in Eq.(\ref{waveisobar}), the integration over $T_{(1,2,3)}$ amplitudes has both branch cut contribution from free three-body Green's function, see in Eq.(\ref{waveisobar}), and poles contribution from scattering amplitudes themselves.  Only branch cut contribution is responsible for diffraction effect, in another word, only branch cut integration creates new final momenta over scattering, pole terms do not create any new momenta.  Branch cut integration is usually troublesome, fortunately,  as we already know from \cite{McGuire:1964zt}, diffraction in total wave function has to be cancelled out. By  some simple algebra in Eq.(\ref{reconwave}), it is easy to see that \mbox{$ \sum_{\gamma=1}^{3} T_{(\gamma)} (k_{\gamma}) \propto (k_{12}^{2}  + \frac{3}{4} k_{3}^{2} -  \sigma^{2}) $}, thus  \mbox{$\frac{\sum_{\gamma=1}^{3} T_{(\gamma)} (k_{\gamma})}{k_{12}^{2}  + \frac{3}{4} k_{3}^{2} -  \sigma^{2}} $} has only pole terms. We first complete the integration of $k_{12}$, and pick up the poles in upper half $k_{12}$ plane for \mbox{$r_{12}>0$}, and the poles in lower half $k_{12}$ plane for \mbox{$r_{12}<0$}, so we get
\begin{widetext}
\begin{align}
& \psi    (r_{12}, r_{3};q_{ij}, q_{k})  =  e^{i q_{12} r_{12}} e^{ i q_{3} r_{3}}  \nonumber   \\
&   +  \frac{  \left (  \frac{ i m V_{0} }{2 q_{23}}  \right )    e^{- i q_{23}  | r_{23} | }e^{i q_{1}  r_{1}  }   }{   1- \frac{ i  m V_{0}}{2 q_{23}}   }      + \frac{  \left ( \frac{ i m V_{0}}{2 q_{31}}    \right ) e^{ - i   q_{31}  | r_{31} | }e^{i q_{2}  r_{2}  }  }{ 1- \frac{ i  m V_{0}}{2 q_{31}}  }    -  \frac{ \left (\frac{ i m V_{0}}{2 q_{12}} \right ) \left [  \left ( \frac{ i  m V_{0}}{2 q_{23}} \right )      e^{  i  q_{12}  | r_{31} | }e^{i q_{3}  r_{2}  }  +e^{  i q_{12}  | r_{12} | }e^{i q_{3}  r_{3}  }  \right ] }{\left (1+ \frac{ i  m V_{0}}{2 q_{12}} \right )\left (1- \frac{ i  m V_{0}}{2 q_{23}} \right )}    \nonumber \\
&     +   \frac{    \left (   i m V_{0} \right )  }{  \left ( 1+ \frac{ i  m V_{0}}{2 q_{12}} \right ) \left (1- \frac{ i  m V_{0}}{2 q_{23}} \right ) \left (1- \frac{ i  m V_{0}}{2 q_{31}} \right )  } i \int_{-\infty}^{\infty}   \frac{d k_{3}}{2\pi}   \nonumber \\
& \quad      \times   \left [ \quad     \frac{ \left (- \frac{ i  m V_{0}}{2 q_{23}} \right )\left (1- \frac{ i  m V_{0}}{2 q_{31}} \right )  \theta(r_{12})  e^{i (\frac{k_{3}}{2}+q_{3}) r_{1 2}}  e^{i k_{3}  r_{3}  }  + \left ( \frac{ i  m V_{0}}{2 q_{31}} \right )\left (1- \frac{ i  m V_{0}}{2 q_{23}} \right )     \theta(-r_{12})  e^{-i (\frac{k_{3}}{2}+q_{3}) r_{1 2}}  e^{i k_{3}  r_{3}  }   }{   \left (k_{3} - q_{2} - i \epsilon  \right) \left  ( k_{3}     - q_{1} + i \epsilon \right )  }  \right. \nonumber \\
& \quad \quad     +   \frac{  \left [ \left ( \frac{ i  m V_{0}}{2 q_{31}} \right )- \left ( \frac{ i  m V_{0}}{2 q_{12}} \right ) \left (1- \frac{ i  m V_{0}}{2 q_{31}} \right )    \right ]  \theta(r_{12})  e^{-i (\frac{k_{3}}{2}+q_{1}) r_{1 2}}  e^{i k_{3}  r_{3}  } +  \left (- \frac{ i  m V_{0}}{2 q_{12}} \right ) \left (1+ \frac{ i  m V_{0}}{2 q_{31}} \right )    \theta(-r_{12})  e^{i (\frac{k_{3}}{2}+q_{1}) r_{1 2}}  e^{i k_{3}  r_{3}  }}{   \left (k_{3} - q_{2} - i \epsilon  \right) \left  ( k_{3}     - q_{3} - i \epsilon \right )  }  \nonumber \\
& \quad \quad     +   \left.  \frac{  \left [ \left ( \frac{ i  m V_{0}}{2 q_{23}} \right )- \left ( \frac{ i  m V_{0}}{2 q_{12}} \right ) \left (1- \frac{ i  m V_{0}}{2 q_{23}} \right )    \right ]  \theta(r_{12})  e^{i (\frac{k_{3}}{2}+q_{2}) r_{1 2}}  e^{i k_{3}  r_{3}  } +  \left ( - \frac{ i  m V_{0}}{2 q_{12}} \right ) \left (1+ \frac{ i  m V_{0}}{2 q_{23}} \right )    \theta(-r_{12})  e^{-i (\frac{k_{3}}{2}+q_{2}) r_{1 2}}  e^{i k_{3}  r_{3}  }  }{   \left (k_{3} - q_{1} + i \epsilon  \right) \left  ( k_{3}     - q_{3} - i \epsilon \right )  }    \right ].
\end{align}
Next, we can perform $k_{3}$ integration and pick up all the poles in both upper and lower half $k_{3}$ plane in a similar manner as we did in $k_{12}$ integration, thus, we finally get 
\begin{align}
 \psi (r_{12},r_{3}; &q_{ij}, q_{k})  =\left ( A e^{i q_{12} r_{12}} + B  e^{ - i q_{12} r_{12}} \right ) e^{i q_{3} r_{3}}    + \left ( C e^{i q_{23} r_{12}} + D e^{ - i q_{23} r_{12}} \right ) e^{i q_{1} r_{3}}  + \left ( E  e^{i q_{31} r_{12}} + F  e^{ - i q_{31} r_{12}} \right ) e^{i q_{2} r_{3}},   \label{wavesolpsi01}
\end{align}
where the coefficients are given by
\begin{align}
A&=1+  \frac{   \theta(- r_{23})  \left (  \frac{ i m V_{0} }{2 q_{23}}  \right )      }{   1- \frac{ i  m V_{0}}{2 q_{23}}   }   + \frac{   \theta(- r_{31})  \left ( \frac{ i m V_{0}}{2 q_{31}}    \right )  }{ 1- \frac{ i  m V_{0}}{2 q_{31}}  } -  \frac{ \theta(r_{12}) \left (\frac{ i m V_{0}}{2 q_{12}} \right ) }{\left (1+ \frac{ i  m V_{0}}{2 q_{12}} \right )\left (1- \frac{ i  m V_{0}}{2 q_{23}} \right )}      \nonumber \\
&+  \theta(r_{12})  \frac{  -  \theta(r_{23}) \left (  \frac{ i  m V_{0}}{2 q_{23}} \right )   \left [ \left ( \frac{ i  m V_{0}}{2 q_{31}} \right )- \left ( \frac{ i  m V_{0}}{2 q_{12}} \right ) \left (1- \frac{ i  m V_{0}}{2 q_{31}} \right )    \right ]     + \theta(-r_{31}) \left ( \frac{ i  m V_{0}}{2 q_{31}} \right ) \left [ \left ( \frac{ i  m V_{0}}{2 q_{23}} \right )- \left ( \frac{ i  m V_{0}}{2 q_{12}} \right ) \left (1- \frac{ i  m V_{0}}{2 q_{23}} \right )    \right ]           }{  \left ( 1+ \frac{ i  m V_{0}}{2 q_{12}} \right ) \left (1- \frac{ i  m V_{0}}{2 q_{23}} \right ) \left (1- \frac{ i  m V_{0}}{2 q_{31}} \right )  } ,  
\end{align}
\begin{align}
B&=-  \frac{ \theta(-r_{12})  \left (\frac{ i m V_{0}}{2 q_{12}} \right ) }{\left (1+ \frac{ i  m V_{0}}{2 q_{12}} \right )\left (1- \frac{ i  m V_{0}}{2 q_{23}} \right )}     +  \theta(-r_{12})   \frac{    \left ( \frac{ i  m V_{0}}{2 q_{12}} \right ) \left [ \theta(-r_{31})   \left ( \frac{ i  m V_{0}}{2 q_{23}} \right )  \left (1+ \frac{ i  m V_{0}}{2 q_{31}} \right )    - \theta(r_{23})    \left ( \frac{ i  m V_{0}}{2 q_{31}} \right )    \left (1+ \frac{ i  m V_{0}}{2 q_{23}} \right )     \right ]}{  \left ( 1+ \frac{ i  m V_{0}}{2 q_{12}} \right ) \left (1- \frac{ i  m V_{0}}{2 q_{23}} \right ) \left (1- \frac{ i  m V_{0}}{2 q_{31}} \right )  } ,  
\end{align}
\begin{align}
C&=-  \frac{\theta(r_{31})    \left (\frac{ i m V_{0}}{2 q_{12}} \right )   \left ( \frac{ i  m V_{0}}{2 q_{23}} \right )    }{\left (1+ \frac{ i  m V_{0}}{2 q_{12}} \right )\left (1- \frac{ i  m V_{0}}{2 q_{23}} \right )}     +   \frac{     \left ( \frac{ i  m V_{0}}{2 q_{12}} \right )  \left [  \theta(r_{12})      \theta(r_{31})   \left ( \frac{ i  m V_{0}}{2 q_{23}} \right )\left (1- \frac{ i  m V_{0}}{2 q_{31}} \right )   -  \theta(-r_{12})    \theta(-r_{23})  \left ( \frac{ i  m V_{0}}{2 q_{31}} \right )    \left (1+ \frac{ i  m V_{0}}{2 q_{23}} \right )   \right ]}{  \left ( 1+ \frac{ i  m V_{0}}{2 q_{12}} \right ) \left (1- \frac{ i  m V_{0}}{2 q_{23}} \right ) \left (1- \frac{ i  m V_{0}}{2 q_{31}} \right )  } ,   
\end{align}
\begin{align}
D&= \frac{ \theta( r_{31}) \left ( \frac{ i m V_{0}}{2 q_{31}}    \right )  }{ 1- \frac{ i  m V_{0}}{2 q_{31}}  }      +  \frac{   \left ( \frac{ i  m V_{0}}{2 q_{31}} \right ) \left \{  -   \theta(-r_{12})      \theta(-r_{23}) \left ( \frac{ i  m V_{0}}{2 q_{12}} \right )\left (1- \frac{ i  m V_{0}}{2 q_{23}} \right )    + \theta(r_{12})   \theta(r_{31})  \left [ \left ( \frac{ i  m V_{0}}{2 q_{23}} \right )- \left ( \frac{ i  m V_{0}}{2 q_{12}} \right ) \left (1- \frac{ i  m V_{0}}{2 q_{23}} \right )    \right ]   \right \} }{  \left ( 1+ \frac{ i  m V_{0}}{2 q_{12}} \right ) \left (1- \frac{ i  m V_{0}}{2 q_{23}} \right ) \left (1- \frac{ i  m V_{0}}{2 q_{31}} \right )  }, 
\end{align}
\begin{align}
E&= - \frac{      \theta(-r_{12})     \theta(r_{23})    \left ( \frac{ i  m V_{0}}{2 q_{12}} \right ) \left ( \frac{ i  m V_{0}}{2 q_{31}} \right )\left (1- \frac{ i  m V_{0}}{2 q_{23}} \right )  +      \theta(-r_{12}) \theta(-r_{31})   \left ( \frac{ i  m V_{0}}{2 q_{23}} \right )   \left ( \frac{ i  m V_{0}}{2 q_{12}} \right ) \left (1+ \frac{ i  m V_{0}}{2 q_{31}} \right )  }{  \left ( 1+ \frac{ i  m V_{0}}{2 q_{12}} \right ) \left (1- \frac{ i  m V_{0}}{2 q_{23}} \right ) \left (1- \frac{ i  m V_{0}}{2 q_{31}} \right )  }  , 
\end{align}
and
\begin{align}
F&= \frac{ \theta(r_{23})  \left (  \frac{ i m V_{0} }{2 q_{23}}  \right )      }{   1- \frac{ i  m V_{0}}{2 q_{23}}   }  -  \frac{ \theta(-r_{31})   \left (\frac{ i m V_{0}}{2 q_{12}} \right )   \left ( \frac{ i  m V_{0}}{2 q_{23}} \right )     }{\left (1+ \frac{ i  m V_{0}}{2 q_{12}} \right )\left (1- \frac{ i  m V_{0}}{2 q_{23}} \right )}    \nonumber \\
& +  \frac{   \theta(r_{12})      \left ( \frac{ i  m V_{0}}{2 q_{23}} \right )  \left \{   \theta(-r_{31}) \left ( \frac{ i  m V_{0}}{2 q_{12}} \right )  \left (1- \frac{ i  m V_{0}}{2 q_{31}} \right )     +   \theta(r_{23})  \left [ \left ( \frac{ i  m V_{0}}{2 q_{31}} \right )- \left ( \frac{ i  m V_{0}}{2 q_{12}} \right ) \left (1- \frac{ i  m V_{0}}{2 q_{31}} \right )    \right ]     \right \} }{  \left ( 1+ \frac{ i  m V_{0}}{2 q_{12}} \right ) \left (1- \frac{ i  m V_{0}}{2 q_{23}} \right ) \left (1- \frac{ i  m V_{0}}{2 q_{31}} \right )  }. \label{coefwavesolpsi01}
\end{align}

\end{widetext}
It can be shown that the coefficients given in Eq.(\ref{coefwavesolpsi01}) are the solutions of McGuire's Model, the coefficients obtained in six individual regions in ($r_{12},r_{3}$) plane, see Fig.~\ref{fig:config}, satisfy matrix transformation conditions in Eq.(\ref{mcguirecond}).

The three-body wave functions for other free incoming waves are obtained in a similar way, because of length expression of these wave functions, we do not show them all in this work except the wave function for three fermions and three bosons system, the expression of   three fermions and three bosons systems are listed in Sections \ref{3fermionsfree} and \ref{3bosonsfree} respectively.

\section{McGuire's Model}\label{McGuire}
\begin{figure}
\begin{center}
\includegraphics[width=3.2 in,angle=0]{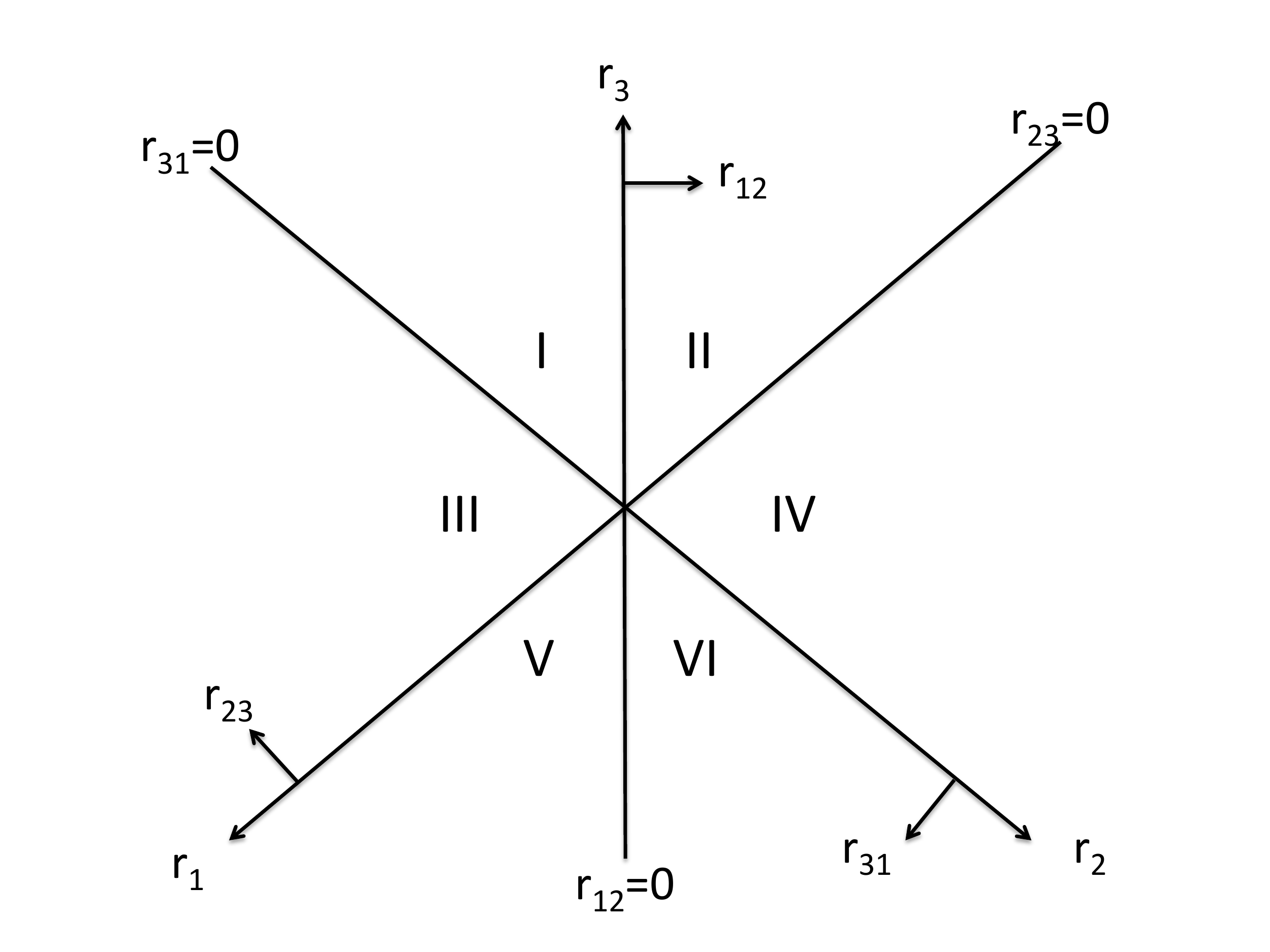}  
\caption{ Diagram show six segments from $(I) - (VI)$ in ($r_{12},r_{3}$) plane, the $\delta$-function potentials are non-zero only at lines, \mbox{$r_{12}=0$}, \mbox{$r_{23}=0$} and \mbox{$r_{31}=0$}.  The arrows show the positive direction of each variable. \label{fig:config}  } 
\end{center}
\end{figure}

One dimensional three identical particles system interacting through equal-strength $\delta$-function potential has been solved by ray-tracing method in \cite{McGuire:1964zt}. After removal of the center-of-mass coordinate, the one dimensional three-body problem resembles the motion of a single particle in a two-dimensional configuration space, {\it e.g.} ($r_{12},r_{3}$) plane. The plane is divided symmetrically into six segments by interaction lines at \mbox{$r_{ij}=0$}  \mbox{$(ij=12,23,31)$}, see Fig.~\ref{fig:config}. According to ray-tracing arguments, author in \cite{McGuire:1964zt} shows that   three-particle only exchange momenta during scattering, no new momenta are generated by collision, hence no diffraction. Therefore a general solution of wave function is a linear combination of six possible plane-waves,
\begin{align}
\psi_{ \Lambda }(r_{12},r_{3})& =\left ( A_{ \Lambda} e^{i q_{12} r_{12}} + B_{ \Lambda} e^{ - i q_{12} r_{12}} \right ) e^{i q_{3} r_{3}}  \nonumber \\
&+ \left ( C_{ \Lambda} e^{i q_{23} r_{12}} + D_{ \Lambda} e^{ - i q_{23} r_{12}} \right ) e^{i q_{1} r_{3}} \nonumber \\
&+ \left ( E_{ \Lambda} e^{i q_{31} r_{12}} + F_{ \Lambda} e^{ - i q_{31} r_{12}} \right ) e^{i q_{2} r_{3}}, \label{mcguirewave}
\end{align}
where $\Lambda $ stands for six segments from \mbox{$(I)$} up to \mbox{$(VI)$}. The  coefficients in six segments are related by boundary conditions of wave function, {\it e.g.} the boundary conditions at \mbox{$r_{12}=0$} between segment $(I)$ and $(II)$ are given by,
\begin{align}
& \psi_{II} (r_{12},r_{3})|_{r_{12} = 0^{+}} =  \psi_{I} (r_{12},r_{3})|_{r_{12} = 0^{-}}, \nonumber \\
& \left. \frac{\partial  \psi_{II} (r_{12},r_{3})}{\partial r_{12}} \right |_{r_{12} = 0^{+}}  - \left. \frac{\partial  \psi_{I} (r_{12},r_{3})}{\partial r_{12}} \right  |_{r_{12} = 0^{-}}   \nonumber \\
& \quad \quad  \quad \quad  \quad \quad  \quad   \quad = m V_{0} \psi_{II} (r_{12},r_{3})|_{r_{12} = 0^{+}}, \label{boundary}
\end{align}
the rest of boundary conditions are given in a similar way. If we define the vector of coefficients by \mbox{$\chi^{T}_{\Lambda}= \left (A_{ \Lambda},B_{ \Lambda}, \cdots ,  F_{ \Lambda} \right )$} in segment $\Lambda$, the two  neighboring $\chi_{\Lambda}$ vectors    are connected by matrix transformation, 
\begin{equation}
\chi_{\Lambda} =\Gamma_{\Lambda, \Lambda' } \chi_{\Lambda'}, \label{mcguirecond}
\end{equation}
 where $\Gamma_{\Lambda, \Lambda' } $ is determined by Eq.(\ref{boundary}).

\begin{widetext}
For completeness, we give the expressions of six $\Gamma $ matrice,
\begin{align} 
  \Gamma_{II,I} 
 =  
  \begin{bmatrix}  
 1 -  \frac{i m V_{0}}{2 q_{12}} &  -  \frac{i m V_{0}}{2 q_{12}}& 0 & 0 & 0 & 0  \\
    \frac{i m V_{0}}{2 q_{12}} & 1+  \frac{i m V_{0}}{2 q_{12}}& 0 & 0 & 0 & 0 \\
0 & 0&  1 -  \frac{i m V_{0}}{2 q_{23}} &  -  \frac{i m V_{0}}{2 q_{23}} & 0 & 0 \\   
0 & 0 & \frac{i m V_{0}}{2 q_{23}} & 1+  \frac{i m V_{0}}{2 q_{23}}  & 0 & 0 \\
0 & 0& 0 & 0&  1 -  \frac{i m V_{0}}{2 q_{31}} &  -  \frac{i m V_{0}}{2 q_{31}}   \\   
0 & 0 & 0& 0& \frac{i m V_{0}}{2 q_{31}} & 1+  \frac{i m V_{0}}{2 q_{31}}
 \end{bmatrix} ,  
\end{align}
\begin{align} 
  \Gamma_{IV,II} 
 =  
 \begin{bmatrix}  
1 + \frac{i m V_{0}}{2 q_{23}} & 0 & 0 & 0 & 0 &  \frac{i m V_{0}}{2 q_{23}}  \\  
 0& 1- \frac{i m V_{0}}{2 q_{31}} &  - \frac{i m V_{0}}{2 q_{31}} & 0 & 0 & 0 \\
0 & \frac{i m V_{0}}{2 q_{31}}&  1 +\frac{i m V_{0}}{2 q_{31}}& 0 & 0 & 0 \\   
0 & 0 &0& 1- \frac{i m V_{0}}{2 q_{12}}  & - \frac{i m V_{0}}{2 q_{12}} & 0 \\
0 & 0& 0 &  \frac{i m V_{0}}{2 q_{12}}&  1+ \frac{i m V_{0}}{2 q_{12}} &   0   \\   
- \frac{i m V_{0}}{2 q_{23}}& 0 & 0& 0& 0& 1- \frac{i m V_{0}}{2 q_{23}}
 \end{bmatrix}  ,  
\end{align}
\begin{align} 
  \Gamma_{VI,IV} 
 =  
\begin{bmatrix}  
1 - \frac{i m V_{0}}{2 q_{31}} & 0 & 0 &  -  \frac{i m V_{0}}{2 q_{31}} & 0 & 0 \\  
0 & 1+\frac{i m V_{0}}{2 q_{23}}& 0 & 0 &  \frac{i m V_{0}}{2 q_{23}}& 0 \\   
0 & 0 &1 -\frac{i m V_{0}}{2 q_{12}}&  0 &  0 & -\frac{i m V_{0}}{2 q_{12}} \\
 \frac{i m V_{0}}{2 q_{31}} & 0 & 0& 1+  \frac{i m V_{0}}{2 q_{31}} & 0 & 0 \\
0 & -\frac{i m V_{0}}{2 q_{23}}& 0 & 0&  1-\frac{i m V_{0}}{2 q_{23}}&   0   \\   
0& 0 &  \frac{i m V_{0}}{2 q_{12}}& 0& 0& 1+\frac{i m V_{0}}{2 q_{12}}
 \end{bmatrix}  ,  
\end{align}
\begin{align} 
  \Gamma_{V,VI} 
 =  
 \begin{bmatrix}  
 1 +  \frac{i m V_{0}}{2 q_{12}} &   \frac{i m V_{0}}{2 q_{12}} & 0 & 0 & 0 & 0  \\
   - \frac{i m V_{0}}{2 q_{12}} & 1-  \frac{i m V_{0}}{2 q_{12}} & 0 & 0 & 0 & 0 \\
0 & 0&  1 + \frac{i m V_{0}}{2 q_{23}}&  \frac{i m V_{0}}{2 q_{23}} & 0 & 0 \\   
0 & 0 &- \frac{i m V_{0}}{2 q_{23}} & 1- \frac{i m V_{0}}{2 q_{23}} & 0 & 0 \\
0 & 0& 0 & 0&  1 + \frac{i m V_{0}}{2 q_{31}} &  \frac{i m V_{0}}{2 q_{31}}    \\   
0 & 0 & 0& 0& -\frac{i m V_{0}}{2 q_{31}}& 1-\frac{i m V_{0}}{2 q_{31}}  
 \end{bmatrix} ,  
\end{align}
\begin{align} 
  \Gamma_{III,V} 
 =  
\begin{bmatrix}  
1 -\frac{i m V_{0}}{2 q_{23}}& 0 & 0 & 0 & 0 &   -\frac{i m V_{0}}{2 q_{23}}  \\  
 0& 1+ \frac{i m V_{0}}{2 q_{31}}&  \frac{i m V_{0}}{2 q_{31}} & 0 & 0 & 0 \\
0 & - \frac{i m V_{0}}{2 q_{31}} &  1 -\frac{i m V_{0}}{2 q_{31}} & 0 & 0 & 0 \\   
0 & 0 &0& 1+\frac{i m V_{0}}{2 q_{12}} & \frac{i m V_{0}}{2 q_{12}}& 0 \\
0 & 0& 0 & - \frac{i m V_{0}}{2 q_{12}} &  1- \frac{i m V_{0}}{2 q_{12}} &   0   \\   
\frac{i m V_{0}}{2 q_{23}}& 0 & 0& 0& 0& 1+ \frac{i m V_{0}}{2 q_{23}} 
 \end{bmatrix} ,  
\end{align}
\begin{align} 
  \Gamma_{I,III} 
 =  
\begin{bmatrix}  
1 -  \frac{i m V_{0}}{2 q_{31}} & 0 & 0 &  - \frac{i m V_{0}}{2 q_{31}}   & 0 & 0 \\  
0 & 1+\frac{i m V_{0}}{2 q_{23}}& 0 & 0 & \frac{i m V_{0}}{2 q_{23}}  & 0 \\   
0 & 0 &1 -\frac{i m V_{0}}{2 q_{12}} &  0 &  0 & -\frac{i m V_{0}}{2 q_{12}} \\
 \frac{i m V_{0}}{2 q_{31}} & 0 & 0& 1+  \frac{i m V_{0}}{2 q_{31}} & 0 & 0 \\
0 & -\frac{i m V_{0}}{2 q_{23}}& 0 & 0&  1-\frac{i m V_{0}}{2 q_{23}}&   0   \\   
0& 0 &  \frac{i m V_{0}}{2 q_{12}}& 0& 0& 1+\frac{i m V_{0}}{2 q_{12}}
 \end{bmatrix} . 
\end{align}

The scattering amplitudes $T_{(\gamma)}$'s can be constructed by using Eq.(\ref{Tgammawav}), therefore, we obtain
\begin{align}
T_{(3)}  ( q;q_{ij}, q_{k}) &=  i m V_{0} \left [  \frac{   A_{ I}  + B_{ I}    }{ q-q_{3} - i \epsilon }    +  \frac{   C_{ I}  + D_{ I}   }{ q-q_{1} - i \epsilon  } +  \frac{  E_{ I}   + F_{ I}     }{q-q_{2} - i \epsilon }  - \frac{  A_{ VI}  + B_{ VI}    }{q-q_{3} + i \epsilon  }  - \frac{  C_{V I}  + D_{ VI}    }{q-q_{1} + i \epsilon } - \frac{  E_{ VI}   + F_{ VI}    }{q-q_{2} + i \epsilon }   \right ] , \nonumber \\
T_{(1)}  ( q;q_{ij}, q_{k}) &=  i m V_{0}  \left [  \frac{   E_{ III}  + D_{ III}    }{ q-q_{3} - i \epsilon }    +  \frac{   A_{ III}  + F_{ III}   }{ q-q_{1} - i \epsilon  } +  \frac{  B_{II I}   + C_{II I}     }{q-q_{2} - i \epsilon }  - \frac{  E_{ II}  + D_{ II}    }{q-q_{3} + i \epsilon  }  - \frac{  A_{I I}  + F_{ I I}    }{q-q_{1} + i \epsilon } - \frac{  B_{ II}   + C_{ II}    }{q-q_{2} + i \epsilon }   \right ], \nonumber \\
T_{(2)}  ( q;q_{ij}, q_{k}) &=  i m V_{0}    \left [  \frac{   C_{ VI}  + F_{ VI}    }{ q-q_{3} - i \epsilon }    +  \frac{   B_{ VI}  + E_{ VI}   }{q- q_{1} - i \epsilon  } +  \frac{  A_{V I}   + D_{V I}     }{q-q_{2} - i \epsilon }  - \frac{  C_{ I}  + F_{ I}    }{q-q_{3} + i \epsilon  }  - \frac{  B_{ I}  + E_{  I}    }{q-q_{1} + i \epsilon } - \frac{  A_{ I}   + D_{ I}    }{q-q_{2} + i \epsilon }   \right ]  . \label{Tsolmcguire}
\end{align}
As we can see, the scattering amplitudes bear no branch cuts, but only pole terms as the consequence of Bethe's hypothesis.

\end{widetext}

\section{Two-body scattering }\label{2bscatt}
For completeness, we also give the brief review of two-body interaction in finite volume in this section.
\subsection{Two-body scattering in free space}
We consider two spinless identical particles scattering, the positions and momenta of two particles are denoted by $(x_{1}, x_{2})$ and $(p_{1}, p_{2})$ respectively. The wave function of scattering two particles satisfies Schr\"odinger  equation,
\begin{align}
\left [- \frac{1}{2 m} \frac{d^{2}}{d x_{1}^{2}}- \frac{1}{2 m} \frac{d^{2}}{d x_{2}^{2}} + V(x_{1}-x_{2}) - E \right ] \Psi(x_{1}, x_{2})=0,
\end{align}
where the mass of particle is $m$,  the total energy of two-particle system is \mbox{$E= \frac{p_{1}^{2}}{2m}+\frac{p_{2}^{2}}{2m}$}. Let's denote the center of mass and relative positions  by \mbox{$R= \frac{ x_{1}+ x_{2}}{2}$}  and \mbox{$r=x_{1} - x_{2}$} respectively, and conjugate momenta by  \mbox{$P = p_{1} + p_{2}$} and \mbox{$k = \frac{p_{1}-p_{2}}{2}$} respectively. Due to   translational invariance of center of mass motion,   the total wave function of two particles is described by the product of a plane wave, $e^{ i P R} $, that describes center of mass motion and the wave function, $\psi(r;k)$, that only describes relative motion of two particles, \mbox{$\Psi(x_{1}, x_{2}) = e^{ i P R} \psi(r;k)$}.
 It may be more convenient to use Lippmann-Schwinger equation representation of  solutions,
\begin{align}
\psi(r;k) &= e^{ i k r}  + \int_{-\infty}^{\infty} d r'  G_{(0)}(r - r'; z_{k}) m V(r') \psi (r';k),
\end{align}
where \mbox{$ z_{k} = k^{2} + i \epsilon $} and  \mbox{$k^{2}= m E -\frac{P^{2}}{4}  $}, the free-particle Green's function is given by
\begin{align}
 G_{(0)}(r ; z_{k})  =  \int_{-\infty}^{\infty}  \frac{d q}{2\pi} \frac{e^{ i q r}}{ z_{k}- q^{2} } = - \frac{i e^{i \sqrt{ k^{2}} |r|}}{2 \sqrt{ k^{2} }}, \label{green2b0}    
\end{align}
 At large separation, \mbox{$|r| \gg |r'|$},  the Green's function can be approximated by
\begin{align}
 G_{(0)}(r - r'; z_{k})   \stackrel{|r| \gg |r'|}{\simeq}  - \frac{i e^{i \sqrt{ k^{2}} |r|}}{2 \sqrt{ k^{2} }}  e^{- i \sqrt{k^{2}} \frac{r r'}{|r|}}.
\end{align}
Therefore,   asymptotically, 
\begin{align}
\psi( r;k) & \stackrel{\mbox{large } |r|}{=}     e^{i k r}   + i  t (k,  k' )  e^{i \sqrt{ k^{2}} |r|} , \label{2bwaveasym}
\end{align}
where \mbox{$k' = \sqrt{ k^{2}} \frac{r}{|r|}$} and the scattering amplitudes are given by
\begin{align}
t(k, k')  =& - \frac{ 1}{2 \sqrt{ k^{2} }}   \int_{-\infty}^{\infty} d r'  e^{- i  k' r'} m V(r') \psi(r';k).
\end{align}
In this work, we only consider particles scattering in a symmetric potential, \mbox{$V(r) = V(-r)$}, therefore, the   Schr\"odinger equation exhibits a solution of even parity (two spinless bosons), \mbox{$\psi_{+}(-r)=\psi_{+}(r)$}, and a solution of odd parity (two spinless fermions),  \mbox{$\psi_{-}(-r)=-\psi_{-}(r)$}, where \mbox{$\psi_{\pm} = \frac{\psi_{k} \pm \psi_{-k}}{2}$}. The parity      amplitudes are given by \mbox{$t(k, k') = t_{+} (\sqrt{ k^{2}}) + \frac{k k'}{ k^{2}} t_{-} (\sqrt{ k^{2}})$}, therefore,
\begin{align}
\psi_{\pm}(r; k) & \stackrel{\mbox{large } |r|}{=}   Y_{\pm}(k ) \nonumber \\
&   \times \left [ \frac{e^{i \sqrt{ k^{2}} r} \pm e^{- i \sqrt{k^{2}} r}}{2} +  i   t_{\pm}   (\sqrt{k^{2}})   e^{i \sqrt{k^{2}} |r|} Y_{\pm}(r )  \right ] , \label{2bwf}
\end{align}
where \mbox{$Y_{+} =1$} and \mbox{$Y_{-} (k  ) =  \frac{k }{\sqrt{k^{2}}} ,  Y_{-} (r  ) =  \frac{r }{|r|}  $}. The general wave function thus is the linear superposition of both parity wave functions:  \mbox{$\psi = c_{+} \psi_{+} + c_{-} \psi_{-}$}.

\subsection{Two-body scattering in finite volume}
When the particles are placed in a one dimensional periodic box with the size of $L$, two-particle wave function in a finite box, \mbox{$\Psi^{(L)}(x_{1},x_{2})$}, has to satisfy periodic boundary condition,
\begin{align}
\Psi^{(L)}(x_{1}+ n_{x_{1}} L,x_{2} + n_{x_{2}} L) = \Psi^{(L)}(x_{1},x_{2}), \ \ n_{x_{1},x_{2}} \in \mathbb{Z}.
\end{align}
The finite volume wave function, $\Psi^{(L)}$,  can be constructed from free space wave function $\Psi$ by
\begin{align}
& \Psi^{(L)}(x_{1},x_{2})  =\frac{1}{V} \sum_{n_{x_{1}}, n_{x_{2}} \in \mathbb{Z}} \Psi(x_{1}+ n_{x_{1}} L,x_{2}+n_{x_{2}} L) \nonumber \\
& \quad \quad \quad \quad \quad    =  \left ( \frac{1}{V} \sum_{n_{x_{1}} \in \mathbb{Z}} e^{i P n_{x_{1}} L}  \right ) e^{i P R} \psi^{(L)} (r;k) ,  \nonumber \\
& \psi^{(L)} (r;k) =  \sum_{n \in \mathbb{Z}} e^{ -i \frac{P  }{2} n L  }  \psi (r+ n L;k),
\end{align}
where  \mbox{$n = n_{x_{1}}-n_{x_{2}}$}, and the volume of infinite summation, $V$, is given by \mbox{$V= \sum_{n \in \mathbb{Z}} e^{i P n L} = \frac{2\pi}{L} \sum_{d \in \mathbb{Z}}  \delta(P + \frac{2 \pi}{L} d) $}.  The quantization of total momentum, \mbox{$P = \frac{2\pi}{L} d$}, is warranted by translational invariance of center of mass motion in a periodic box. By our construction,   the general relative wave function in finite box   is given by \mbox{$\psi^{(L)} = c_{+} \psi^{(L)}_{+} + c_{-} \psi^{(L)}_{-}$}, the periodic boundary condition for $\psi^{(L)}$ reads
\begin{align}
 \psi^{(L)} (r+ n L ; k) = e^{ i \frac{P  }{2} n L  } \psi^{(L)} (r ; k). 
\end{align}
Applying Eq.(\ref{2bwf}), the relative wave functions in finite box, $\psi^{(L)}_{\pm}(r; k)$, are given by
\begin{align}
& \psi^{(L)}_{\pm}(r; k)   \stackrel{\mbox{large } |r|}{=}    i    t_{\pm}   (\sqrt{k^{2}}) Y_{\pm} (k) \sum_{n \in \mathbb{Z}} e^{- i \frac{P  }{2} n L  }  \nonumber \\
 & \quad \times  \left [ \pm \theta(-r- n L)   e^{- i \sqrt{k^{2}} (r+ n L)} + \theta(r+ n L)   e^{ i \sqrt{k^{2}} (r+ n L)} \right ].
\end{align}
The summations can be carried out,
\begin{align}
 & \sum_{n \in \mathbb{Z}} e^{ -i \frac{P  }{2} n L  }   \theta(-r- n L)   e^{- i \sqrt{k^{2}}  n L}  \nonumber \\
 & \quad \quad  \quad \quad=\theta(-r)  +  \frac{  e^{i (\frac{P}{2} + \sqrt{k^{2}}) L } }{1- e^{i (\frac{P}{2} + \sqrt{k^{2}}) L }}  ,  \nonumber \\
   & \sum_{n \in \mathbb{Z}} e^{ -i \frac{P  }{2} n L  }  \theta(r+ n L)   e^{ i \sqrt{k^{2}}  n L}  \nonumber \\
 & \quad \quad  \quad \quad = \theta(r)  + \frac{  e^{i (-\frac{P}{2} + \sqrt{k^{2}}) L } }{1- e^{i (-\frac{P}{2} + \sqrt{k^{2}}) L }} ,  
\end{align}
therefore, we find
\begin{align}
  \psi^{(L)}_{\pm} & (r; k)   \stackrel{\mbox{large } |r|}{=}    i    t_{\pm}   (\sqrt{k^{2}}) Y_{\pm} (k)  \nonumber \\
 &   \times  \bigg [ e^{i \sqrt{k^{2}} |r|} Y_{\pm} (r)  +  \frac{  e^{i (-\frac{P}{2} + \sqrt{k^{2}}) L } }{1- e^{i (-\frac{P}{2} + \sqrt{k^{2}}) L }}  e^{i \sqrt{k^{2}} r}  \nonumber \\
 &   \quad \quad   \quad \quad   \quad \quad    \pm    \frac{  e^{i (\frac{P}{2} + \sqrt{k^{2}}) L } }{1- e^{i (\frac{P}{2} + \sqrt{k^{2}}) L }}    e^{-i \sqrt{k^{2}} r}   \bigg ].
\end{align}

Secular equation is obtained by matching $\psi^{(L)}(r)$ to $\psi(r)$ at an arbitrary $r$, larger than the range of the interaction. The matching procedure is equivalent to applying periodic condition to both wave functions and derivative of wave functions at nearest neighbor when solving periodic potential quantum mechanics problems.   Because wave functions are the linear superposition of two independent basis, $e^{\pm i \sqrt{k^{2}} r}$, by choosing \mbox{$r>0$} {\it e.g.}, we obtain two    matching equations,
\begin{align}
& \left [ \frac{1}{2 i t_{+}} -  \frac{ e^{i (-\frac{P}{2} + \sqrt{k^{2}}) L } }{1- e^{i (-\frac{P}{2} + \sqrt{k^{2}}) L }}  \right ] c_{+}  \nonumber \\
& \quad \quad +  \left [ \frac{1}{2 i t_{-}}  - \frac{ e^{i (-\frac{P}{2} + \sqrt{k^{2}}) L } }{1- e^{i (-\frac{P}{2} + \sqrt{k^{2}}) L }} \right ] c_{-} =0, \nonumber \\
& \left [ \frac{1}{2 i t_{+}} - \frac{   e^{i (\frac{P}{2} + \sqrt{k^{2}}) L } }{1- e^{i (\frac{P}{2} + \sqrt{k^{2}}) L }} \right ] c_{+}  \nonumber \\
& \quad \quad - \left [ \frac{1}{2 i t_{-}} - \frac{   e^{i (\frac{P}{2} + \sqrt{k^{2}}) L } }{1- e^{i (\frac{P}{2} + \sqrt{k^{2}}) L }}  \right ] c_{-}=0.
\end{align}
The above equations have non-trivial solutions when
\begin{align}
&  1 - (2 i t_{+}  + 2 i t_{-} ) \frac{1}{2} \left [ \frac{ e^{i (\frac{P}{2} + \sqrt{k^{2}}) L } }{1- e^{i (\frac{P}{2} + \sqrt{k^{2}}) L }}   +  \frac{ e^{i (-\frac{P}{2} + \sqrt{k^{2}}) L } }{1- e^{i (-\frac{P}{2} + \sqrt{k62}) L }}   \right ] \nonumber \\
&+2 i t_{+}  2 i t_{-}   \frac{ e^{i (\frac{P}{2} + \sqrt{k^{2}}) L } }{1- e^{i (\frac{P}{2} + \sqrt{k^{2}}) L }}     \frac{ e^{i (-\frac{P}{2} + \sqrt{k^{2}}) L } }{1- e^{i (-\frac{P}{2} + \sqrt{k^{2}}) L }}  =0. \label{2bdet}
\end{align}
Due to \mbox{$e^{i P L}=1$}, it is clearly to see that  the solutions of secular equation, Eq.(\ref{2bdet}), can be divided into classes of   positive parity state solutions and negative parity state solutions, the definite parity state solutions are given by equation
\begin{equation}
e^{-i (\frac{P}{2} + \sqrt{k^{2}}) L } = 1 + 2 i t_{\mathcal{P}} (\sqrt{k^{2}}), \ \ \ \ \mathcal{P} = \pm. \label{2bpritdet}
\end{equation}
When scattering amplitudes are parameterized by phase shifts, \mbox{$t_{\pm} = \frac{1}{\cot \delta_{\pm}-i}$}, the secular equations, Eq.(\ref{2bdet}) and Eq.(\ref{2bpritdet}), are reduced respectively to non-relativistic version of L\"uscher's formula in one dimension \cite{Guo:2013vsa},
\begin{align}
& \cos \frac{P L}{2} = \frac{\cos (\delta_{+} + \delta_{-} + \sqrt{k^{2}} L)}{\cos (\delta_{+} - \delta_{-})} , \\
& \cot \delta_{\mathcal{P}} + \cot \frac{\frac{PL}{2} + \sqrt{k^{2}} L}{2} =0. 
\end{align}
In following subsections, we show the recovery of analytic solutions  for two well-known one dimensional models by applying quantization condition obtained in Eq.(\ref{2bpritdet}).

\subsection{Solvable examples of two-body scattering in finite volume}

\subsubsection{Kronig Penney model}
Let's consider square well potential \mbox{$V(r)=V_{0}$} for \mbox{$|r|< \frac{b}{2}$}, and \mbox{$V(r)=0$} otherwise. The symmetric wave functions in short range,  \mbox{$|r|< \frac{b}{2}$}, are given by,
\begin{align}
\psi_{\pm}(r; k)  =  A_{\pm}    \frac{e^{i \sqrt{\sigma^{2}_{V}} r} \pm e^{- i \sqrt{\sigma^{2}_{V}} r}}{2} ,  \ \ \ \ |r|< \frac{b}{2},
\end{align}
where \mbox{$\sigma^{2}_{V} = k^{2} - m V_{0} $}, continuity of wave functions at boundary of potential leads to relations,
\begin{align}
& A_{\pm} =\frac{2 e^{- i \sqrt{k^{2}}  \frac{b}{2} } }{ \pm \left (1- \frac{\sqrt{\sigma^{2}_{V}}}{\sqrt{k^{2}}}  \right ) e^{ i \sqrt{\sigma^{2}_{V}}  \frac{b}{2} }  +  \left (1+ \frac{\sqrt{\sigma^{2}_{V}}}{\sqrt{k^{2}}}  \right ) e^{- i \sqrt{\sigma^{2}_{V}}  \frac{b}{2} } } , \nonumber \\
& 1+ 2 i t_{\pm}  = e^{- i \sqrt{k^{2}}  b }  \frac{\cos \sqrt{\sigma^{2}_{V} } \frac{b}{2}  + \left ( \frac{\sqrt{\sigma^{2}_{V}}}{\sqrt{k^{2}}}  \right)^{\pm 1} i \sin \sqrt{\sigma^{2}_{V}  } \frac{b}{2} }{\cos \sqrt{\sigma^{2}_{V}  } \frac{b}{2}  - \left ( \frac{\sqrt{\sigma^{2}_{V}}}{\sqrt{k^{2}}}  \right)^{\pm 1} i \sin \sqrt{\sigma^{2}_{V} } \frac{b}{2} } . \label{kronig}
\end{align}
Easy to check, the scattering amplitudes $t_{\pm} $ are also the solutions of
\begin{equation}
 t_{\pm} = - \frac{ 1}{2 \sqrt{k^{2}}}   \int_{-\frac{b}{2}}^{\frac{b}{2}} d r'   e^{- i  \sqrt{k^{2}}  r'}  m V_{0} \psi_{\pm} (r'; k).
\end{equation}

In finite box with periodic boundary condition, plugging the analytic expression of $t_{\pm}$ in Eq.(\ref{kronig}) into the secular equation Eq.(\ref{2bdet}),  we thus obtain well-known energy quantization condition for Kronig Penney model,
\begin{align}
  \cos \sqrt{k^{2} } a  \cos  \sqrt{\sigma^{2}_{V}} b &  -  \frac{k^{2} + \sigma^{2}_{V}}{2 \sqrt{k^{2}} \sqrt{\sigma^{2}_{V}} } \sin \sqrt{k^{2} } a  \sin  \sqrt{\sigma^{2}_{V}} b   \nonumber \\
&           = \cos \frac{PL}{2},   \ \ \ \ \ \  a = L-b.
\end{align}

\subsubsection{$\delta$-function potential model}
Now, let's consider a short range interaction model with a delta potential, \mbox{$V(r)=V_{0} \delta(r)$}, the amplitudes for $\delta$-function potential thus are given by
\begin{equation}
 t_{\pm} (\sqrt{k^{2}})= - \frac{ 1}{2 \sqrt{k^{2}}}     m V_{0} \psi_{\pm} (0), 
\end{equation}
where \mbox{$\psi_{+}(0) =1+ i t_{+}$} and \mbox{$\psi_{-}(0)=0$}. Therefore, we obtain,
\begin{equation}
it_{+}  (\sqrt{k^{2}})=-  \frac{ \frac{i m V_{0}}{2 \sqrt{k^{2}}} }{1+ \frac{ i m V_{0}}{2 \sqrt{k^{2}}}}, \ \ \ \  i t_{-}=0.  \label{deltatamp}
\end{equation}
Plugging the solution of $i t_{+}$ into secular equation, Eq.(\ref{2bdet}), thus, we obtain the well-known  quantization condition for two-particle interaction in a finite box with a periodic boundary condition,
\begin{equation}
e^{-i (\frac{P}{2} + \sqrt{k^{2}}) L } =   \frac{ 1- \frac{ i m V_{0}}{2 \sqrt{k^{2}}}}{1+ \frac{ i m V_{0}}{2 \sqrt{k^{2}}}}.
\end{equation}
The results of delta potential can also be obtained from Kronig Penny model by taking the limit of \mbox{$b\rightarrow 0$}, \mbox{$V_{0}\rightarrow \infty$} and \mbox{$b V_{0} = \mbox{const}$}.

\end{document}